%% file: main.tex
\documentclass[review]{elsarticle}

\usepackage{lineno,hyperref}
\usepackage{amsmath}
\usepackage{xcolor}

\journal{Journal of \LaTeX\ Templates}

\begin{document}

\begin{frontmatter}

\title{Level-set topology optimization considering nonlinear thermoelasticity}

\author[unist]{Hayoung Chung}

\author[technion]{Oded Amir}
\ead{odedamir@technion.ac.il}

\author[ucsd,cardiff]{H. Alicia Kim\corref{mycorrespondingauthor}}
\cortext[mycorrespondingauthor]{Corresponding author}
\ead{alicia@ucsd.edu}

\address[unist]{School of Mechanical, Aerospace and Nuclear Engineering, Ulsan National Institute of Science and Technology, Ulsan, 44919, Republic of Korea}
\address[technion]{Civil and Environmental Engineering, Technion, Technion City, Haifa, 32000, Israel}
\address[ucsd]{Structural Engineering Department, University of California, San Diego, 9500 Gilman Drive, San Diego, CA 92093, USA}
\address[cardiff]{Cardiff School of Engineering, Cardiff University, The Queen’s Buildings, 14-17 The Parade, Cardiff, CF24 3AA, United Kingdom }

\begin{abstract}
At elevated temperature environments, elastic structures experience a change of the stress-free state of the body that can strongly influence the optimal topology of the structure. 
This work presents level-set based topology optimization of structures undergoing large deformations due to thermal and mechanical loads. 
The nonlinear analysis model is constructed by multiplicatively decomposing thermal and mechanical effects and introducing an intermediate stress-free state between the undeformed and deformed coordinates. 
By incorporating the thermoelastic nonlinearity into the level-set topology optimization scheme, wider design spaces can be explored with the consideration of both mechanical and thermal loads. 
Four numerical examples are presented that demonstrate how temperature changes affect the optimal design of large-deforming structures.
In particular, we show how optimization can manipulate the material layout in order to create a counteracting effect between thermal and mechanical loads, even up to a degree that buckling and snap-through are suppressed.  
Hence the consideration of large deformations in conjunction with thermoelasticity opens many new possibilities for controlling and manipulating the thermo-mechanical response via topology optimization.

\end{abstract}


\end{frontmatter}


\input{includes/intro_jun24}

\input{includes/FE}

\input{includes/TopOpt}

\input{includes/Results_new}

\input{includes/Conclusion_new}

\section{Acknowledgement}
Chung and Kim acknowledge the support from DARPA (Award number HR0011-16-2-0032) and NASA’s Transformation Tools and Technologies Project (grant number 80NSSC18M0153). Kim also acknowledges the support of the Engineering and Physical Sciences Research Council Fellowship for Growth (grant number EP/M002322/2). The authors thank the Numerical Analysis Group at the Rutherford Appleton Laboratory for their FORTRAN HSL packages (HSL, a collection of Fortran codes for large-scale scientific computation).

\section{Conflict of Interest}
On behalf of all authors, the corresponding author states that there is no conflict of interest. 
\section*{References}

\bibliography{mybibfile}

\end{document}

%% file: includes/intro_jun24.tex
\section{Introduction}

Thermoelasticity broadly refers to a coupled phenomenon where the elastic responses of a structure are affected by a temperature change. 
The phenomenon is widely considered in diverse engineering disciplines in association with various structural responses, ranging from elastic behaviors such as classical volumetric expansion to the deterioration of the structural integrity such as thermal buckling, delamination, and fracture. 
Recently, thermoelastic behaviors have also been employed to realize non-conventional behavior of materials, e.g., metamaterials \cite{qu2017micro, wu2016isotropic}, and phase-changing smart materials \cite{yuan2017shape, de2014programmed}. 

Topology optimization of structures experiencing thermoelastic load was first studied by Rodrigues and Fernandes \cite{Rodrigues1995}, where the classical compliance objective was extended to accommodate thermal load combined with mechanical loads. 
Therein, a design scheme that follows the material distribution approach with homogenization was presented.
Such a coupled load was later considered also with other approaches, such as Evolutionary Structural Optimization, or ESO \cite{Li2001}; and the level-set method \cite{Xia2008}.
By solving the thermo-compliance minimization problem, these schemes demonstrated how optimized structures accommodate a uniform temperature change. 
Due to its simplicity and analogy with the classical structural compliance minimization problem, thermo-compliance minimization has been widely adopted in later works, including design that involves multi-material \cite{Gao2016} and coupled physics \cite{de2007topological}. 

To further discuss structural resistance to thermoelastic loads, Pedersen and Pedersen \cite{Pedersen2010} presented topology optimization considering mechanical strength, where they identified that different ratios between mechanical and thermal loads lead to different designs.   
In the same line of thought, Deaton and Grandhi \cite{Deaton2013} examined the difference between optimized layouts that are derived based on the different definitions of the compliance. 
Therein thermal and mechanical compliances are defined as $\boldsymbol{u}^{thT} \boldsymbol{f}^{th}$, and $\boldsymbol{u}^{mT} \boldsymbol{f}^m$, where $\boldsymbol{u}$ and $\boldsymbol{f}$ represent the nodal displacement and the load vector, respectively. 
Superscripts $th$ and $m$ indicate that the terms are either thermal or mechanical. 
Comparative studies were also conducted by Zhang et al., \cite{Zhang2014}, where the topological layouts resulting either from strain energy minimization or mean compliance minimization are compared and discussed. 
These lead to a search of alternative thermoelastic merit functions, such as a constraint on local stress concentration \cite{Takalloozadeh2017, Neiferd2018}, and a thermoelastic buckling constraint \cite{deng2017topology}.

Multifunctional actuators and metamaterials, 
that are designed to exhibit a thermally driven shape change by 
meticulously distributing active materials, are another class of applications of topology optimization considering thermoelasticity. 
In the seminal works by Sigmund and Torquato \cite{sigmund1997design}, and Sigmund \cite{sigmund2001designI, sigmund2001designII}, metamaterials of negative thermal expansion, and micro-electromechanical actuators are designed, respectively. 
The idea of thermally driven actuation and multiphysical coupling has been widely explored thereafter, leading to the works including the extensions to different topology optimization approaches \cite{luo2009shape, luo2012meshfree}, the designs of piezo-electric actuators \cite{ruiz2018optimal, ruiz2018optimal_math, kogl2005topology, carbonari2005design}, and the designs of recent 4D printed structures that exhibit self-morphing behavior \cite{maute2015level, geiss2019combined}.

Structures may experience thermoelastic loads that induce a large deflection of the structure, e.g., kinetic heating in aerostructures \cite{Deaton2015}. 
In such cases, nonlinear thermoelasticity must be considered, meaning that the infinitesimal strain should be replaced with nonlinear strain; 
otherwise, displacement and stress states obtained by structural analysis are highly overestimated due to the neglected stress-stiffening effect, and \color{black} these inaccurate variables \color{black} potentially lead to non-optimal structures \cite{chung2019nonlinear}. 
Such an inaccuracy is exacerbated when temperature change is involved \cite{Deaton2015, chung2015light} as the thermoelastic behavior incorporates the temperature-induced deformation that in effect changes the stress-free state with respect to the undeformed state \cite{Lubarda2004}. 
\color{black}
Therefore, it has been concluded that incorporation of the nonlinearity in the thermoelastic topology optimization is significant in getting the optimal design when mechanical and thermal loads are applied to the structure \cite{Deaton2015}.

\color{black}
Consideration of geometric nonlinearity in the thermoelastic design is first presented by Jog \cite{jog1996distributed}. 
The study showed that the nonlinearity does influence the material distribution; meanwhile, the scope of the results is limited as the nonlinearity is mild and the resulting layout is populated with the intermediate densities. 
\color{black}
The nonlinear method is later extended to design large-deforming compliant mechanisms with only thermal or electrothermal loads \cite{sigmund2001designI, sigmund2001designII}, which makes the structures deform without a mechanical load. 
Therein, the effect of the nonlinearity in the multiphysics actuations has been clearly demonstrated by comparing the shapes and the optimalities of these designs with the corresponding linear solutions. 
%
\color{black} 



\color{black} 
When a mechanical load is simultaneously imposed with a thermal load, and potentially induces a large deformation in the design problem, the investigation of the combined load becomes more interesting. 
It is because of the contradicting effect between mechanical and thermal load, meaning that decreasing the material volume during optimization increases the influence of the fixed mechanical load, while decreases that of thermal load \cite{Zhang2014}.
Therefore, the optimized material layouts are expected to consider such a contradicting effect of the two loads. 
In this work, we present a level-set based topology optimization framework that is able to design thermoelastic structures experiencing both large mechanical and temperature loads, while attaining crisp solid-void topologies.
\color{black}

In this respect, the present work is also in line with the early works on the optimization with nonlinear elasticity under mechanical loading only, which was first considered by Buhl et al. \cite{Buhl2000}. 
The resulting material layouts are shown to deviate from the linear solution as the external load increases. 
A difference between the compliance and complementary work objectives is also evident and the corresponding layout changes are discussed.  
Jung and Gea \cite{Jung2004} further discussed the layout change by examining the separate influence of material and geometric nonlinearities on the optimum layouts.
A recent work by Li et al., \cite{li2018shape} further expands the usage of nonlinear elasticity in topology optimization to design that preserves local geometric features constraining warpage.
Allaire \cite{Allaire2004} and Kwak and Cho \cite{Kwak2005} extended nonlinear topology optimization by using a level-set method. 
Recently, optimum layouts obtained by the nonlinear elastic level-set topology optimization were discussed also in terms of attenuation of local buckling \cite{Chen2017}.

Numerous works were dedicated to topological design of structures devoid of structural instabilities.
Bruns et al. \cite{Bruns2002} proposed a combination of the arc-length scheme with the classical Newton-Raphson solver, in order to robustly converge to a structure that undergoes snap-through. 
Kemmler \cite{Kemmler2005} identified the snap-through phenomenon and loss of stability of the intermediate layouts found during the optimization.
These understandings are later applied to optimizing buckling stiffness \cite{lindgaard2013compliance, Jansen2014}, and designing snapping mechanisms \cite{james2016layout}.
Recently, Wallin et al. \cite{wallin2018stiffness} addressed the structural stability of the optimum layouts depending on the choice of objective function that in effect optimizes different types of stiffness. 

In this work, we formulate and develop nonlinear thermoelastic level-set topology optimization.
\color{black} 
In particular, we investigate the intertwined effects of the mechanical and thermal loads on optimized large-deforming structural designs. 
\color{black}
Both geometric and material nonlinearities are considered in the analysis and sensitivity calculation. 
A thermally induced change of the natural state, which significantly affects the stress state, is identified by the multiplicative decomposition of the strain tensor \cite{belytschko2013nonlinear, Lubarda2004, geiss2019combined}.
These nonlinear thermoelastic considerations enhance the accuracy in estimating internal states for higher loading conditions, hence extend the range of both mechanical and thermal loads to the extent where the structure deforms by the same order as its own characteristic dimension. 
The temperature range, for example, roughly coincides with what aero-structures experience under high-temperature operating conditions \cite{Deaton2015}. 
To demonstrate such nonlinear effects, bending-dominant and buckling-dominant structures are examined as a basis for discussing how temperature changes affect the optimum designs. 

The remainder of the paper is organized as follows. 
Section 2 describes the thermoelastic formulation and its corresponding finite element analysis model. 
The Total Lagrangian approach is implemented based on the deformation gradient that is multiplicatively decomposed into its mechanical and thermal parts. 
The optimization problem and the corresponding shape sensitivity used in the level-set topology optimization are presented in Section 3. 
In Section 4 we present four demonstrative examples of which material layouts are discussed in detail, and investigate the combined effect of the nonlinearity and the thermal loads. 
Finally, the conclusion is given in Section 5.  

%% file: includes/FE.tex
\section{Nonlinear Thermoelastic Finite Element Model}

In this section, the finite element model for the nonlinear thermoelastic problem used in this study is outlined. 
First of all, nonlinear kinematics of the thermoelastic deformation is presented.
For describing large deformations, multiplicative decomposition is used and the consistent Green-Lagrange strain is formulated. 
Then the finite element model that uses a standard Galerkin method is briefly discussed, where the equilibrium equations in the discretized functional space and the iterative solvers are shown.
Lastly, the modified hyperelastic model, which accounts not only for a constitutive relation but also for preventing distorted elements, is addressed.

\subsection{Kinematics of the thermoelastic structure}

\begin{figure}[hbt!]
    \centering
    \includegraphics[width=10cm]{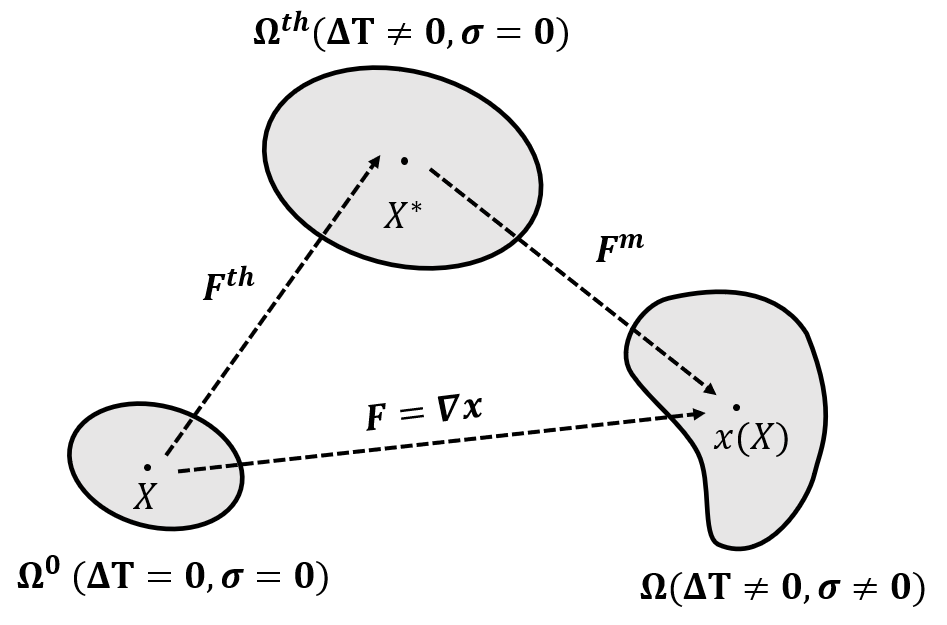}
    \caption{Kinematic representation of the deforming body. The intermediate configuration $\Omega^{th}$ is introduced between the deformed $\Omega$ and the undeformed $\Omega^0$ ones}
    \label{fig:Decomposition}
\end{figure}

Kinematics of the thermoelastic large deformation is briefly introduced in the context of finite elasticity.
To understand responses of a structure experiencing thermoelastic loads, the displacement of the structure is decomposed into purely elastic and thermal parts.
A multiplicative decomposition of the deformation gradient $\boldsymbol F$ is adopted herein,  
which is a general approach when inelastic deformation is incorporated within the structural behavior \cite{belytschko2013nonlinear, Holzapfel2002}. 
Such a decomposition method is also addressed in \cite{Lubarda2004,Darijani2013,Holzapfel2002} in the context of reversible thermoelastic deformation.
Figure \ref{fig:Decomposition} illustrates the material configurations of the deforming body and the associated displacements.

Between the deformed ($\Omega$) and undeformed configuration ($\Omega^0$), the intermediate material configuration $\Omega^{th}$ is introduced.
This is a virtual state that is conceptually introduced by isothermal elastic destressing of the deformed configuration to a stress-free state ($\sigma=0$). 
\color{black} 
A deformation gradient $\boldsymbol F = \nabla_{\boldsymbol X} \boldsymbol{x}$, which linearly maps the displacement of the initial material point $\boldsymbol{X} \in \Omega^0$ to $\boldsymbol{x} \in \Omega$, 
is thereby decomposed as follows:
\color{black}
\begin{equation}
\label{eqn:F_decompose}
\boldsymbol F = \boldsymbol F^m \cdot \boldsymbol F^{th} 
\end{equation} 
where $\boldsymbol F^{th}$ and $\boldsymbol F^m$ each refers to thermal and mechanical part, respectively. 
The change of the metric between $\Omega^0$ to $\Omega^{th}$ is associated with $\boldsymbol F^{th} = \nabla_{\boldsymbol{X}} \boldsymbol{X^*}$, whereas the isotropic elastic deformation between $\Omega^{th}$ to $\Omega$ is represented by $\boldsymbol F^m = \nabla_{\boldsymbol X^*} \boldsymbol x $. 
The increase of the strain energy, and the elastic constitutive relation are assumed to be dependent on $\boldsymbol F^m$ only. 
This formulation accounts for the origin of the thermally-induced mechanical stress.
Given that there is no observable deformation (i.e. $\boldsymbol{F=I}$) in the elevated temperature condition $\boldsymbol{F^{th}} \neq \boldsymbol{I}$, the mechanical energy changes due to the mechanical deformation, which is calculated via $\boldsymbol{F}^m =\boldsymbol{F}\cdot\boldsymbol{F}^{th-1} \neq \boldsymbol{I}$.
\color{black}
Without loss of generality, the elastic Green-Lagrange strain $\boldsymbol E^m$ is computed by, 
\begin{equation}
    \begin{aligned}
        2 \boldsymbol E^m 
        & = (\boldsymbol F^m)^T \boldsymbol F^m \\
        & = (\boldsymbol F^{th})^{-T} (\boldsymbol F^T \boldsymbol F) (\boldsymbol F^{th})^{-1} \\
        & = 2(\boldsymbol F^{th})^{-T} (\boldsymbol E-\boldsymbol E^{th}) (\boldsymbol F^{th})^{-1} 
        \label{eqn:multiplicativeE}
    \end{aligned}
\end{equation}
where the thermal strain $\boldsymbol E^{th} = \frac{1}{2} ((\boldsymbol F^{th})^T \boldsymbol F^{th} - \boldsymbol I)$. 

The thermally induced gradient $\boldsymbol F^{th}$ is assumed to be dependent upon the temperature change $\Delta T$ and thermal expansion coefficient $\alpha$. When the thermal expansion is isotropic and linear with respect to a temperature change, $\boldsymbol F^{th}$ is specified by 
\begin{equation}	
    \boldsymbol F^{th} = (1 + \alpha \Delta T) \boldsymbol I
\label{eqn:isotropicTemp}
\end{equation} 
where $\boldsymbol I$ refers to an identity tensor. 

By the isotropy assumption (\ref{eqn:isotropicTemp}) the Green-Lagrange strain calculation in (\ref{eqn:multiplicativeE}) reduces the additive decomposition of elastic strain and thermal strain.
This decouples the strains and simplifies the finite element formulation. 
However, for generality, we employ the multiplicative decomposition in this paper as such an equivalence is not valid in anisotropic non-elastic expansion.
The discussion regarding additive decomposition can be found in \cite{Darijani2013}.

\subsection{Discretized force equilibrium}

The finite element formulation used herein employs the standard Galerkin method and Total Lagrangian approach. For details, the readers are referred to the textbooks on nonlinear finite element methods and continuum mechanics \cite{belytschko2013nonlinear, Holzapfel2002}.

Assuming that the body force is negligible, the equilibrium equation becomes 
\begin{equation}
    \begin{aligned}
        &\nabla (\boldsymbol P)= \nabla (\boldsymbol S \boldsymbol F^T) = 0  \quad \text{in} \quad \Omega^0 \\
        &\boldsymbol{P(F)\cdot n} = t \quad  \text{in}  \quad \partial\Omega^0_N        
    \end{aligned}
\label{eqn:staticEquil}
\end{equation}
where $\boldsymbol{P}$ and $\boldsymbol{S}$ denote the first and second Piola-Kirchhoff total stress, respectively, evaluated in the undeformed configuration. 
$\boldsymbol{n}$ is a normal vector of the Neumann boundary $\partial \Omega_N^0$, and $t$ is a surface traction.
\color{black}
$\boldsymbol P$ is computed as the derivative of the strain energy $W$ with respect to its work conjugate $\boldsymbol F$,
\begin{equation}
    \boldsymbol P = \frac{\partial W}{\partial \boldsymbol F}
    = \frac{\partial W}{\partial \boldsymbol F^m}\frac{\partial \boldsymbol F^m}{\partial \boldsymbol F} 
    =  (\boldsymbol F^{th})^{-1} \cdot \boldsymbol P^m 
\label{eqn:1stPK_scaled}
\end{equation}	
where $\boldsymbol P^m$ refers to the mechanical part of $\boldsymbol P$. 

Isoparametric polynomial shape function $N$ is employed for the finite element discretization, i.e., $u_i = N(\boldsymbol X)^I u_i^I$. 
\color{black} The \color{black} capitalized superscripts indicate the index for node, whereas the subscripts denotes a degree of freedom. 
Based on Eqn. (\ref{eqn:F_decompose}) and (\ref{eqn:1stPK_scaled}), the internal force vector $\boldsymbol{f}^{int}$ of a given temperature change $\Delta T$ and the displacement $\boldsymbol u$ is derived: 
\begin{equation}
    \begin{aligned}
        f^{int,I}_i (\boldsymbol u,\Delta T) &= \int_{\Omega^0} B^I_j S_{jk} (\boldsymbol u, \Delta T) F_{ik} (u) 
        \label{eqn:FE_fint}
    \end{aligned}
\end{equation}	
where $B^I_j$ indicates the gradient of the shape function, i.e. $B^I_j = \nabla_{X_j} N^I(\boldsymbol X)$.

By perturbing terms found in Eqn.~(\ref{eqn:FE_fint}) by $\boldsymbol u$, the tangent stiffness $K^{IJ}_{ij}$ as a summation of the material term $K^{IJ,M}_{ij}$ and the geometric term $K^{IJ,G}_{ij}$ is derived:
\begin{equation}
\begin{aligned}
    K^{IJ,M}_{ij} &= \int_{\Omega^0} \bigg(B^I_k F^{th-1}_{kl} F^{m}_{im}\bigg) C^{SE}_{lmdb}  
                                     \bigg(B^J_f F^{th-1}_{fd} F^{m}_{jb}\bigg)  det(F^{th})^{-1} \\
    K^{IJ,G}_{ij} &= \int_{\Omega^0} \bigg(B^I_k F^{th-1}_{kl}\bigg) S^m_{lm} 
                                     \bigg(B^J_b F^{th-1}_{bm} \bigg)  \delta_{ij} det(F^{th})^{-1}
    \label{eqn:FE_K}
    \end{aligned}
\end{equation}	
where the Einstein summation notation is used for both subscript and superscript. 
\color{black}
$\boldsymbol C^{SE}$ is the Hessian of the strain energy $W$ with respect to the Green-Lagrange strain $E$. 
\color{black}

Based on Eqns.~(\ref{eqn:FE_fint})-(\ref{eqn:FE_K}),
the zero-residual condition at the equilibrium $\boldsymbol r(\boldsymbol u,\Delta T) = \boldsymbol f^{int} - \boldsymbol f^m = 0$ is obtained by the conventional Newton-Raphson algorithm. 
Without loss of generality, either a fixed external load (load-control) scheme or a displacement (displacement-control) scheme is adopted in this work. 
The choice of the control parameter depends on the stability of the intermediate solutions, which are obtained during optimization. 
If the quadratic convergence of the Newton-Raphson solver is maintained at every iteration, a load-control scheme is used for simplicity.
Otherwise, a displacement-control scheme is employed in calculating the static equilibrium.
The solution space is assumed to contain multiple configurations with 
structural instability.
In both cases, the size of the increment parameter is adaptively controlled.
First, the increment size is changed based on the powered ratio between the number of iterations required at the previous increment step and the desired number of iterations \cite{bathe1983our, batoz1979incremental}, which is set to 4 in this study. 
Additionally, the increment size is reduced by a half whenever a convergence is not achieved within the maximum number of iterations, 
which typically originates from an ill-conditioned tangent stiffness matrix. 
Throughout the iterative search, $\Delta T$ is assumed to be constant as for both the the displacement and load control schemes.

\subsection{Hyperelastic constitutive model}

A hyperelastic material is employed to model the nonlinear constitutive relation. The stress-strain constitutive relationship is derived based on the strain energy function $W$ which is independent of strain rate and its history. In this work, a neo-Hookean model for a compressible material is employed: 
\begin{equation}
W = \int_{\Omega^0} \frac{\lambda}{2}(ln J)^2 + \frac{\mu}{2}(tr(\boldsymbol C)-3) - \mu (ln J)
\end{equation}
where $\boldsymbol C$ is a Green deformation tensor, i.e., $\boldsymbol C=\boldsymbol F^T \boldsymbol F$, and $J$ refers to the Jacobian of the deformation gradient. $\lambda$ and $\mu$ are Lam{\'e} constants in the limit of small strains. For simplicity, we set ($\lambda, \mu$) to (1.0, 0.4).

The ersatz material model \cite{Allaire2004} is used to represent the structural layout of $\Omega^0$, 
where the cut elements are represented by a fraction of the material volume within the element. 
When the design domain is discretized by $N_e$ finite elements, the corresponding energy is calculated by 
\begin{equation}
W = \sum_e^{N_e} W_e(\boldsymbol F^m) \rho_e
\label{eqn:discreteW}
\end{equation}
where $\rho_e$ and $W_e$ indicate a material density and strain energy found in a cut element $e$, respectively. 
The numerical singularity of the tangent stiffness matrix in (\ref{eqn:FE_K}) originating from zero-material elements is avoided by imposing fictitious weak material that has $10^{-6}$ modulus of the solid material. 
The elastic deformation gradient and energy are evaluated at the Gauss points within each element. 

The cut elements with the weaker material stiffness are populated near the boundary of a structure, and these elements often distort excessively. 
This phenomenon can cause difficulties in convergence during the search of an elastic equilibrium. 
Numerical manipulations are thereby required to regularize the excessive mesh distortion. 
In this work, the linear interpolation method suggested by Wang et al. \cite{Wang2014} is implemented, which adds a fictitious linear strain energy to near-void elements. 
The interpolation method replaces the strain energy of the element by adding an auxiliary strain energy $W^L$ depending on the parameterized material density $\gamma_e$ found in the element. 
For a given displacement in an element $\boldsymbol{u_e}$, the original hyperelastic energy $W_e$ of element $e$ is replaced by $\hat{W_e}$: 
\begin{equation}
    \begin{aligned}
    \hat{W_e}(\boldsymbol{F}(\boldsymbol{I}+\nabla\boldsymbol{u_e})) &= W_e(\gamma_e \boldsymbol{u_e}) - W^L(\gamma_e \boldsymbol{u_e}) + W^L(\boldsymbol{u_e}) \\
    W^L &= \frac{1}{2} \boldsymbol{\epsilon^L} : \boldsymbol {C^L} : \boldsymbol {\epsilon^L}
    \end{aligned}
\end{equation}
where $\boldsymbol{\epsilon^L}$ is linear strain computed as $(\nabla\boldsymbol{u_e}+\nabla\boldsymbol{u_e}^T)/2$, and $\boldsymbol C^L$ is a stiffness tensor computed based on prescribed Lam{\'e} constants, $\lambda$ and $\mu$. 
Notably, $\hat{W}_e$ is still a function of displacement $\boldsymbol{u}_e$ and ersatz material density $\rho$, as $W_e$ is likewise.
Therefore, the accompanied changes in computing the derivatives due to the modification of the strain energy is straightforward. 
\color{black}
$\gamma_e$ parameterizes the projected density $\rho$ via a smooth Heaviside function:
\begin{equation}
    \gamma = \frac{tanh(\beta \rho_0) + tanh (\beta (\rho -\rho_0))}{ tanh(\beta \rho_0) + tanh (\beta (1-\rho_0))}
\end{equation}
where the smoothing parameters $\beta$ and $\rho_0$ are set to be 500, and 0.01, respectively. 
The interpolated strain energy $\hat{W_e}$ becomes the original hyperelastic energy $W_e (\boldsymbol{u_e})$ when $\gamma_e=1$ and the linear energy $W^L(\boldsymbol{u_e})$ when $\gamma_e=0$. 
This method adds a fictitious energy to the structure and changes the deformation found in low-density materials. 
However, the effects on the overall optimum layouts is observed to be negligible \cite{Wang2014}.
\color{black}

%% file: includes/TopOpt.tex
\section{Topology Optimization}

In this section, the topology optimization problems and their design sensitivities are presented. 
First the compliance minimization problem is redefined in the context of thermoelastic nonlinear response. 
The level-set topology optimization method is then introduced, followed by a derivation of the shape sensitivities defined at the boundary of the domain.

\subsection{Problem formulation} 
The problem is formulated to minimize the thermoelastic compliance for a specified volume constraint. 
The definition of the thermoelastic compliance is not unique \cite{Zhang2014, Takalloozadeh2017, Xia2008, Deaton2013} as the response of the structures are attributed to either temperature change or mechanical loading. 
In this work, we optimize a compliance measure that comprises of the total displacement and the mechanical loading at the final equilibrium point. 
In the context of nonlinear elasticity without thermal loads, this is typically referred to as the end-compliance \cite{Buhl2000}.
When the load-control scheme is used, the objective can be interpreted as minimizing the observable displacement of the structure at the point of the applied mechanical loading, 
\color{black}
\begin{equation}
  \begin{aligned}
    \text{minimize } &\boldsymbol u^T \boldsymbol f^m \\
    \text{subject to } &\|\Omega\| \leq \|\Omega^*\| \\
    & \boldsymbol{r(u)} = \boldsymbol{f}^{int}(\boldsymbol{u})-\boldsymbol{f^m} = \boldsymbol{0}
    \label{eqn:compliance_loadCtrl}
  \end{aligned}
\end{equation}
where $\| \Omega^*\|$ refers to the material volume constraint, $\boldsymbol r$ is a residual between internal $\boldsymbol{f}^{int}$ and external load $\boldsymbol{f}^m$. 
\color{black}
As the objective function depends only on the overall displacement obtained at the end of the iteration of the Newton solver, the optimized design is expected to have maximum secant stiffness 
\color{black}
of the global load-displacement curve \cite{wallin2018stiffness}. 
In the thermoelastic context, the given objective essentially penalizes the displacement to the mechanical direction by leveraging a thermally induced shape change;
depending on the material layout that changes the way the structure deforms by thermal expansion, the structure can deflect in the opposite direction of the mechanical load, hence the objective function can be negative in the contrary to the mechanical-only case. 
As a result, although the objective function found in Eqn.~\eqref{eqn:compliance_loadCtrl} is still to be referred to as the end-compliance, it is not the mechanically stiffest structure that is optimal in the current problem formulation. 
\color{black}

As noted by Wallin et al.~\cite{wallin2018stiffness}, the minimization of the end-compliance as in (\ref{eqn:compliance_loadCtrl}) essentially considers the secant stiffness hence unstable intermediate equilibrium points may exist within the solution path. 
In such cases, it can be difficult or even impossible to find a viable equilibrium at the specified load magnitude $\boldsymbol f^m$ using a load-controlled scheme. 
In such cases, displacement control should be used and the problem formulation is altered accordingly. 
The optimum structure is now expected to sustain the largest force $\boldsymbol f^m$ for the given controlled displacement $u_p$: \color{black}
\begin{equation}
  \begin{aligned}
    \text{minimize } &-\boldsymbol u_{ctrl}^T \boldsymbol f^m = - \theta(\boldsymbol u_{ctrl}^T \hat{\boldsymbol f^{ref}}) \\
    \text{subject to } &\|\Omega\| \leq \|\Omega^*\| \\
    & \boldsymbol{r(u)} = \boldsymbol{f}^{int}(\boldsymbol{u})-\theta \hat{\boldsymbol{f}^{ref}} = \boldsymbol{0}
  \label{eqn:compliance_dispCtrl}
  \end{aligned}
\end{equation}
where the load $\boldsymbol{f}^{m}$ is computed based on the scalar load multiplication factor $\theta$, which is now a state variable. 
The reference load vector $\hat{\boldsymbol f^{ref}}$ is constant throughout the analysis. 
While $u_p$ is specified at a controlled node, $\hat{\boldsymbol f^{ref}}$ can be distributed to multiple nodes.
\color{black}
For convenience in notation, constant controlled displacement vector $\boldsymbol u_{ctrl}$ has only one nonzero value $u_p$ at the controlled node.
This objective, Eqn.~\eqref{eqn:compliance_dispCtrl} was recently used in the context of nonlinear elasticity to achieve buckling-resistant topological layouts of trusses and frame structures with imperfections \cite{madah2017truss,madah2019concurrent}.

\subsection{Level-set topology optimization}

Level-set topology optimization (LSTO) is used to solve the problems shown in Eqn.~(\ref{eqn:compliance_loadCtrl}) and (\ref{eqn:compliance_dispCtrl}).
The topology of a structure $\Omega^0$ is represented by its enclosing boundary $\partial \Omega^0$, which is defined as a zero hypersurface of the signed distance function $\phi(\boldsymbol x)$ on the design domain, i.e., $\Omega^0 = \{x:\phi(\boldsymbol x) \geq 0\}$.
The structure is updated by solving the Hamilton-Jacobi equation, 

\begin{equation} 
\dot{\phi} + \nabla \phi \cdot \frac{\partial\boldsymbol x}{\partial t} 
= \dot{\phi} + V_n \|\nabla\phi\| = 0 
\label{eqn:H-J}
\end{equation}
where $V_n$ is the advection velocity normal to the boundary, and $t$ is the pseudo time step of the advection problem. 
Due to the implicit boundary representation and update scheme, a smooth and well-defined boundary is guaranteed during topological changes, such as a merge of the voids. 
\color{black}
The level-set topology optimization \cite{dunning2015introducing, Sivapuram2016} does not require a filtering scheme or a length scale control. 
Nevertheless, the characteristic size of the structural features found in the optimized design is limited with the size of the element as it is not possible to represent many boundaries cutting a level set element.  
\color{black}
The parameter $V_n$ which is critical in updating the layout towards its optimum, is determined by solving a linearized optimization subproblem. 
In this paper, the numerical scheme is briefly outlined. Interested readers are referred to Picelli et al. \cite{Picelli2018_cmame} and Sivapuram et al. \cite{Sivapuram2016} for further details. 

A topology optimization problem is stated as, 
\begin{equation}
  \begin{gathered}
    \text{minimize } c(\Omega, u) \\ 
    \text{subject to } g(\Omega, u) \leq 0
    \label{eqn:optimProb}  
  \end{gathered}  
\end{equation}

The change of the objective function $c$ and constraint function $g$ can be written as 
\begin{equation}
\Delta \{c,g\} = \bigg\{ \frac{\partial c}{\partial \Omega}, \frac{\partial g}{\partial \Omega} \bigg\}\Delta \Omega
\label{eqn:linearOptim}
\end{equation}
after linearizing (\ref{eqn:optimProb}) with respect to $\Omega$. The boundary is then discretized by $B$ number of line segments and (\ref{eqn:linearOptim}) is reduced to
\begin{equation}
  \begin{gathered}
  \frac{\partial c}{\partial \Omega}\cdot  \Delta \Omega = \Delta t \sum_j^B{V_n^j S_c^j l^j} \\
  \frac{\partial g}{\partial \Omega} \cdot  \Delta \Omega = \Delta t \sum_j^B{V_n^j S_g^j l^j}\\
  \label{eqn:discretizedOptim}
  \end{gathered}
\end{equation}
where $S_c^j$ and $S_g^j$ refer to the design sensitivities evaluated at the discretized boundary point, and $l$ is a length segment found in the boundary point whose index is $j$. $V_n^j \Delta t$ is assumed to be $\alpha d^j$, where $\alpha $ is a distance that the boundary travels along the unit direction $d$. According to Ref.~\cite{Picelli2018_cmame}, (\ref{eqn:discretizedOptim}) is further reduced to searching an optimal $\alpha $ and $\lambda$ as follows,

\begin{equation}
  \begin{aligned}
    \text{minimize }& \Delta t \sum_j^B \bigg\{S_c^j  l^j V_n^j(\alpha, \lambda)\bigg\} \\
    \text{subject to }& \Delta t \sum_j^B \bigg\{S_g^j  l^j V_n^j(\alpha, \lambda)\bigg\} \leq -\bar{g} \\
    & z^j_{min} \leq \alpha d^j \leq z^j_{max} \\
    \text{where } & d^j = \frac{l^j S_c^j + \lambda l^j S_g^j}{\| l^j S_c^j + \lambda l^j S_g^j \|} 
  \label{eqn:suboptim_end}
  \end{aligned}
\end{equation}
where $\bar{g}$ is the change of the constraint function, and $z_{min}$ and $z_{max}$ are limits of the movement, i.e. $V_n \Delta t$, that is calculated based on the geometric consideration of the boundary movement limitation and the CFL (\color{black}Courant--Friedrichs--Lewy\color{black}) condition.

\subsection{Shape sensitivity}
As evident from (\ref{eqn:suboptim_end}), the consistent design sensitivities $S_c$ and  $S_g$ are important parameters in solving the optimization problem.
We follow the variational approach of boundary perturbation suggested by Allaire \cite{Allaire2004} to calculate the consistent shape sensitivities.

For the minimization problems (\ref{eqn:compliance_loadCtrl}, \ref{eqn:compliance_dispCtrl}), an objective $c$ and material volume constraint $g$ functions are generalized to 
\begin{equation}
  \begin{gathered}
  c(\Omega^0, \boldsymbol u) = \int_{\partial \Omega^0} l(u) \\
  g(\Omega^0) = \int_{\Omega^0} H(\phi) 
  \label{eqn:generalProblem}
  \end{gathered}
\end{equation}
where $l(u)$ stands for the generalized compliance computation regardless of the displacement or load control scheme, and $H(\boldsymbol x)$ is a Heaviside function. 
The augmented Lagrangian $L$ considering equation \eqref{eqn:generalProblem} and the static equilibrium equation \eqref{eqn:staticEquil} is proposed as following:
\begin{equation}
  \begin{aligned}
    L(\Omega^0, \boldsymbol{u}, \boldsymbol{q}) = &c(\Omega^0,\boldsymbol{u})+\int_{\Omega^0}\boldsymbol{P(F)}:\nabla\boldsymbol{q} - \int_{\partial\Omega^0_N} \boldsymbol{t}\cdot\boldsymbol{q} \\
    &- \int_{\partial\Omega^0_D}\boldsymbol{q}\cdot \boldsymbol{P(F)}\cdot \boldsymbol{n} + \boldsymbol{u}\cdot \boldsymbol{P(I+\nabla q)}\cdot \boldsymbol{n}    
  \end{aligned}
\end{equation}
where $\boldsymbol{q}$ is an adjoint parameter. 
The Lagrangian is equivalent to the objective function in the static equilibrium, and the shape sensitivity is derived by evaluating a stationary point with respect to $\Omega^0$.
The partial derivative of $L$ with respect to $\Omega^0$ to the direction $\vartheta$ is:
\begin{equation}
  \begin{aligned}
    <\frac{\partial L}{\partial \Omega^0} (\omega, \boldsymbol{u}, \boldsymbol{q}), \vartheta> = & 
    \int_{\partial\Omega^0} (\boldsymbol{P(F)}:\nabla\boldsymbol{q}) \vartheta\cdot \boldsymbol{n} 
    + \int_{\partial\Omega^0} (\frac{\partial l}{\partial \boldsymbol{n}}+\kappa l(\boldsymbol{u})) \vartheta\cdot \boldsymbol{n}  \\
    &- \int_{\partial\Omega^0_N} (\frac{\partial (\boldsymbol{t\cdot q})}{\partial \boldsymbol{n}}+\kappa (\boldsymbol{t\cdot q})) \vartheta\cdot \boldsymbol{n} \\
    & - \int_{\partial\Omega^0_D} (\frac{\partial (\boldsymbol{u\cdot P(I+\nabla q)}+\boldsymbol{q\cdot P(F)}\cdot \boldsymbol{n})}{\partial \boldsymbol{n}}) \vartheta\cdot \boldsymbol{n} 
  \end{aligned}
  \label{eqn:shapesensitivity}
\end{equation}
where $\kappa$ refers a curvature of the boundary. 
\color{black}
A detailed procedure to calculate the shape derivatives can be found in Ref. \cite{Allaire2004}.

Equation \eqref{eqn:shapesensitivity} is further reduced by fixing a boundary (i.e., $\vartheta = 0$) where the nonhomogeneous Neumann boundary is prescribed. 
We also assume that the advecting distance of the boundary in the normal direction (i.e. $\vartheta \cdot n$) as a descent direction $-V_n$, which assures a decrease of the objective function during iterative updates:
\color{black}
\begin{equation}
\begin{aligned}
  c^\prime &= -\int_{\partial \Omega^0} (P:\nabla q) V_n = \int_{\partial \Omega^0} S_c V_n\\
    g^\prime &= -\int_{\partial \Omega^0} H(\phi) V_n = \int_{\partial \Omega^0} S_g V_n
\end{aligned}
\end{equation}
where $q$ is an adjoint parameter calculated via solving a set of linear adjoint equations, which is derived by stationary condition of $L$ with respect to $\boldsymbol{u}$: 
\begin{equation}
  \begin{aligned}
    \nabla \cdot (\boldsymbol{C^{SE}}\nabla \boldsymbol{q})=0 \quad \quad &\text{in} \quad \quad \Omega^0 \\
    \boldsymbol{q} = \boldsymbol{0} \quad \quad &\text{on} \quad \quad \partial\Omega_D^0 \\
    (\boldsymbol{C^{SE}\nabla q})\cdot\boldsymbol{n} = -\frac{\partial l(\boldsymbol{u})}{\partial \boldsymbol{u}} \quad \quad & \text{on} \quad \quad \partial \Omega^0_N     
  \end{aligned}
  \label{eqn:adjointEqns}
\end{equation}
where $\partial \Omega_D$ and $\partial \Omega_N$ each refers to Dirichlet and Neumann boundary,
\color{black} 
and $\boldsymbol{C^{SE}}$ is a Neo-Hookean material stiffness tensor. 
\color{black} 

In contrast to the classical linear compliance minimization that is self-adjoint, two additional computational steps must be added. 
The first is to evaluate the effective adjoint force based on the definition $l^\prime (\boldsymbol u)$ in (\ref{eqn:generalProblem}). 
Another is to compute the consistent adjoint based on (\ref{eqn:adjointEqns}) at equilibrium. 
When the load-control scheme is employed, the computational cost of these additional steps are negligible as the same pre-factorized tangent stiffness matrix $\boldsymbol K_T$ is reused after the equilibrium is searched \cite{Allaire2004}. 
In the displacement-control scheme, a slight modification must be accompanied as one displacement variable is fixed. 
\color{black} 
The coupled system of adjoint equations becomes:
\begin{equation}
  \begin{bmatrix}
    \boldsymbol{K_T} & \hat{\boldsymbol{f}^{ref}} \\
    \hat{\boldsymbol{f}^{ref}}^T & \boldsymbol{0} 
  \end{bmatrix}
  \begin{pmatrix}
    \boldsymbol{q} \\ \boldsymbol{\xi}
  \end{pmatrix}
  =
  \begin{pmatrix}
    \boldsymbol{0} \\ \boldsymbol{-u}_{ctrl}^T\hat{\boldsymbol{f}^{ref}}
  \end{pmatrix}
\end{equation}
where $\xi$ is a Lagrange multiplier to impose a constraint equation $\boldsymbol{f}^{ref}\cdot \boldsymbol{q} = -\boldsymbol{u}_{ctrl}^T \hat{\boldsymbol{f}^{ref}}$.
If a fixed mechanical reference load $\hat{\boldsymbol{f}^{ref}}$ is applied to a single degree of freedom where $u_p$ is applied to, the adjoint parameter $\boldsymbol{q}$ has only non-zero element $-u_p$ at the displacement-prescribed degree of freedom. 
\color{black} 
Further details regarding the adjoint sensitivity in the displacement-control scheme can be found in Ref.~\cite{amir2011efficient}. 
\color{black}

\input{includes/NR_detail}

%% file: includes/NR_detail.tex
\subsection{Computational implementation}
\begin{figure}[hbt!]
    \centering
    \includegraphics[width=12cm]{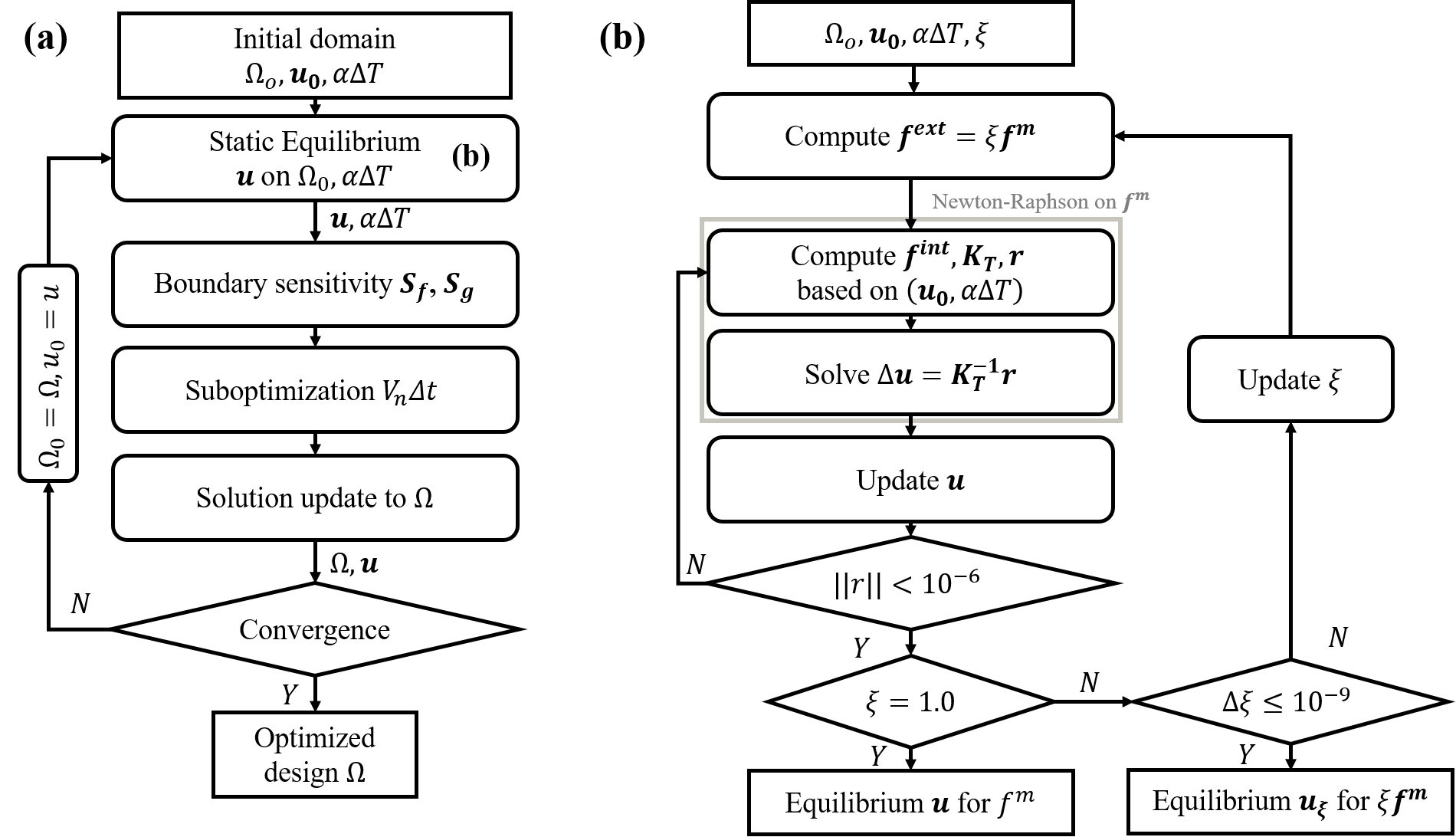}
    \caption{Diagram of the optimization scheme with load-control. (a) Overall scheme where finite element analysis and level-set topology optimization are presented; (b) Detailed procedure for computing the equilibrium at a given load $f^m$.}
    \label{fig:OptScheme}
\end{figure}

The Hamilton-Jacobi equation \eqref{eqn:H-J} and the consistent sensitivities based on the static equilibrium (\ref{eqn:staticEquil}) are computed in every iterations. 
The overall optimization algorithm is summarized in Fig. \ref{fig:OptScheme}. 

Fig. \ref{fig:OptScheme}(a) illustrates the overall nonlinear optimization scheme used herein.
At the beginning of the optimization, an initial design $\Omega_0$ and an assumed equilibrium state $u_0$ are specified. 
The static equilibrium of the structure is searched through nonlinear finite element analysis as presented in Section 2. 
The convergence of the design is checked after every update, i.e., 
the relative difference of the objective values in the last 5 iterations is less than $10^{-4}$, and the constraint are satisfied. 
For computing the optimal boundary movement vector $V_n\Delta t$ \eqref{eqn:suboptim_end}, we employ an SLSQP (sequential least-squares quadratic programming) algorithm in the NLopt package \cite{johnson2014nlopt}. 

Fig.~\ref{fig:OptScheme}(b) presents the numerical scheme used to determine an equilibrium.
We primarily adopted the Newton-Raphson solver that determines the equilibrium incrementally at increased load parameters $\xi \in [0, 1]$. 
The tangent stiffness $K_T$ computed based on Eqn.~\ref{eqn:FE_K} is utilized to find $\boldsymbol u$ that induces zero residual $\boldsymbol r(\boldsymbol u, \alpha \Delta T) = \mathbf{0}$. 
The equilibrium is assumed to be achieved when $\|\boldsymbol r\| \leq 10^{-6}$. 
However, as noted by several authors \cite{Wang2014, Kemmler2005, Bruns2002, Buhl2000}, computation of the equilibrium is not always straightforward in the nonlinear elastic problem.
The main challenges are an efficiency of the computation, and non-feasible solution for a given load. 

We adopted two techniques that alleviate these problems. 
First of all, we reuse a displacement vector $\boldsymbol{u}$, from the previous design iteration, by setting it as the initial guess $\boldsymbol{u_0}$ with $\xi=1$. 
Such a reusing technique is widely employed in topology optimization to expedite a search of the static equilibrium.
In general it effectively reduces the number of iterations, partly because of the gradual update of the material layout 
thanks to the advection method based on the Hamilton-Jacobi equation. 
Although it may negate some of the detailed responses (i.e. new bifurcation point), we found its effect on optimization was negligible. 
However, we noted that there are some cases where the static equilibrium of the current design is largely deviated from the initial guess $\boldsymbol{u_0}$, causing the solution to diverge. In these cases, $\boldsymbol{u_0}$ is reset to zero vector and a typical Newton iteration is restarted.

The second technique applies to when $\Delta \xi \leq 10^{-9}$, which indicates that it is not possible to increase the load parameter $\xi$ for some reason, such as a newly induced structural instability. 
In such cases, we stop the solution search and set the last displacement vector $\boldsymbol{u}_\xi$ as an equilibrium for the given $\xi \boldsymbol f^m$. 
Henceforth, the computation of the boundary sensitivity is based on an intermediate equilibrium, which is used in updating the solution. 
As the optimization workflow is not halted even in the case when the true equilibrium is not attained, the present remedy helps continuing iterations seamlessly. 
An occurrence of such events is found to be correlated with the convergence speed during optimization, which is partly controlled by the CFL condition.  
In this work, we set the CFL condition to be 0.3, so that in the numerical experiments the number of occurrences of intermediate equilibrium was lower or equal to 5. 
Considering the overall optimization typically converged after  200 to 300 iterations, the event is still considered to be rare. 

%% file: includes/Results_new.tex
\section{Numerical Results}

This section presents four examples that clearly demonstrate how the thermal effect influences the optimum solutions when nonlinearity is considered.
The first three examples are statically indeterminate. 
They are affected by the boundary conditions that constrain free thermal expansion; 
reaction forces that are created by such constraints are attained throughout optimization.
The first two examples show a bending-dominant configuration, while the last two examples consider a compressed beam.  
In these examples, geometric nonlinearity of a structure that can consider a large rigid body rotation and structural instability (e.g. snap-through) is significant. 
In all cases, the isolated influences of the nonlinearity due to the mechanical load on the optimized layouts are first discussed. 
\color{black}
These designs are optimized for mechanical loads of various sizes and directions 
\color{black}
and are shown to accommodate these changes by changing a distribution of the materials, which is in contrast to the linear-based designs. 
These results serve as the reference for the subsequent optimization with additional thermal loads. The thermal effect is then examined by varying the combination of temperature and mechanical loads. 
The solutions demonstrate the intricate effect of the temperature change that goes beyond results in the literature that are limited to small displacements.

\subsection{Bi-clamped beam}
\begin{figure}[hbt!]
    \centering
    \includegraphics[width=12cm]{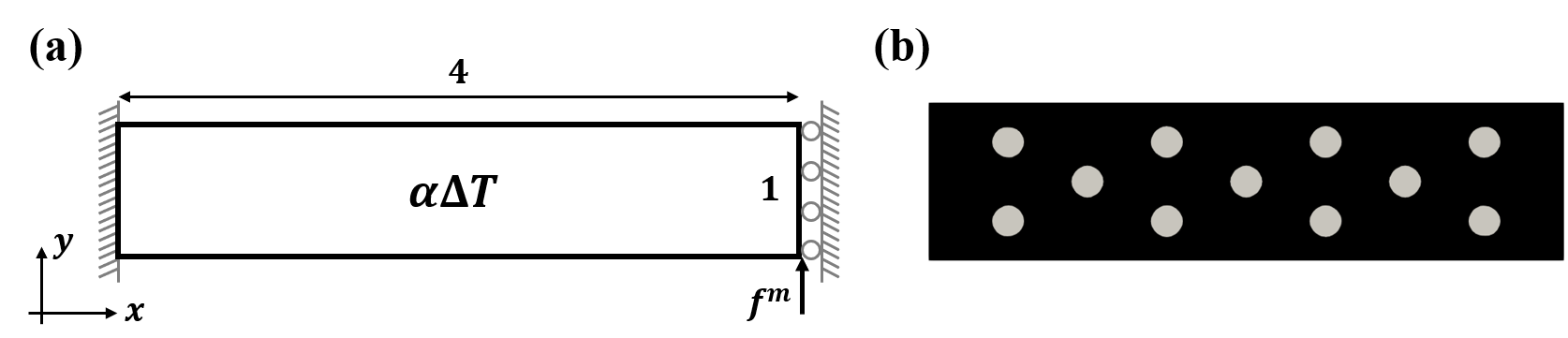}
    \caption{Design configuration of the bi-clamped beam. (a) Dimensions and boundary conditions of the problem; (b) Initial material distribution, where the void material is marked by gray}
    \label{fig:Res1_config}
\end{figure}

The first example considers a bi-clamped beam with a concentrated load at the center of its lower side, Fig. \ref{fig:Res1_config}(a). 
The change of temperature, $\Delta T$, is assumed to be distributed within the structure.
Hence reaction forces are created at the clamped boundaries where the temperature-induced volumetric changes are inhibited. 
Being a statically indeterminate structure, the current problem configuration is widely used, typically with the uniform temperature assumption \cite{Zhang2014, Xia2008, Deaton2013, Pedersen2010}.
This problem is widely used in investigating structural design considering thermoelasticity \cite{Xia2008, Zhang2014, Takalloozadeh2017}. 
However, only linear elasticity has been considered therein,  which limits the design capability of the methods within a small range of both thermal expansions and mechanical loads. 
To increase the nonlinear effect, a beam of higher aspect ratio that reduces the effective bending rigidity is adopted as shown in Ref.~\cite{Jung2004}, where the geometric nonlinearity is considered for the case of mechanical loading only. 
The problem specification and the initial structure are shown in Fig.~\ref{fig:Res1_config}. 

Due to the symmetry, only the left half of the structure is modeled and designed to expedite \color{black}the search of the solution. \color{black}
However, it is widely known that exploiting the symmetry inhibits asymmetric buckling modes from being accounted for during optimization 
hence potentially prevents the design from converging to the actual optimum \cite{lindgaard2013compliance}.
To consider the effect of exploiting symmetry, we conducted optimization without the symmetry, i.e. modeling the full $8\times1$ bi-clamped beam for the selected thermoelastic loads.
\color{black} Such inspection about the applied symmetry is executed \color{black}whenever the optimized layout obtained by symmetric topology optimization is expected to experience the structural instability. 
Although not shown herein, the resulting optimized layouts, obtained with or without symmetry, agree well. 
The effect is therefore found to be negligible for the specific bi-clamped configuration considered herein.
The design domain is discretized by 6400 quadrilateral elements, and the volume limit is set to 30\% of the total domain. 
The initial material layout is shown in the Fig. \ref{fig:Res1_config}(b), where the void area (gray) is contrasted to the material region (black). 

\subsubsection{Mechanical loading only}
A mechanical design problem without temperature change is solved first.
The magnitude of the mechanical loads varies from $10^{-7}$ to $10^{-2}$, to the extent where the maximum displacement of the  structure exceeds at least 50\% of the y-dimension of the domain. 
To prevent excessive mesh distortions in the loaded elements, the nodal force is distributed to 5 neighboring nodes. 
These nodes are considered to be non-designable, i.e.~the level-set cannot move at these nodes.

\begin{figure}[hbt!]
    \centering
    \includegraphics[width=12cm]{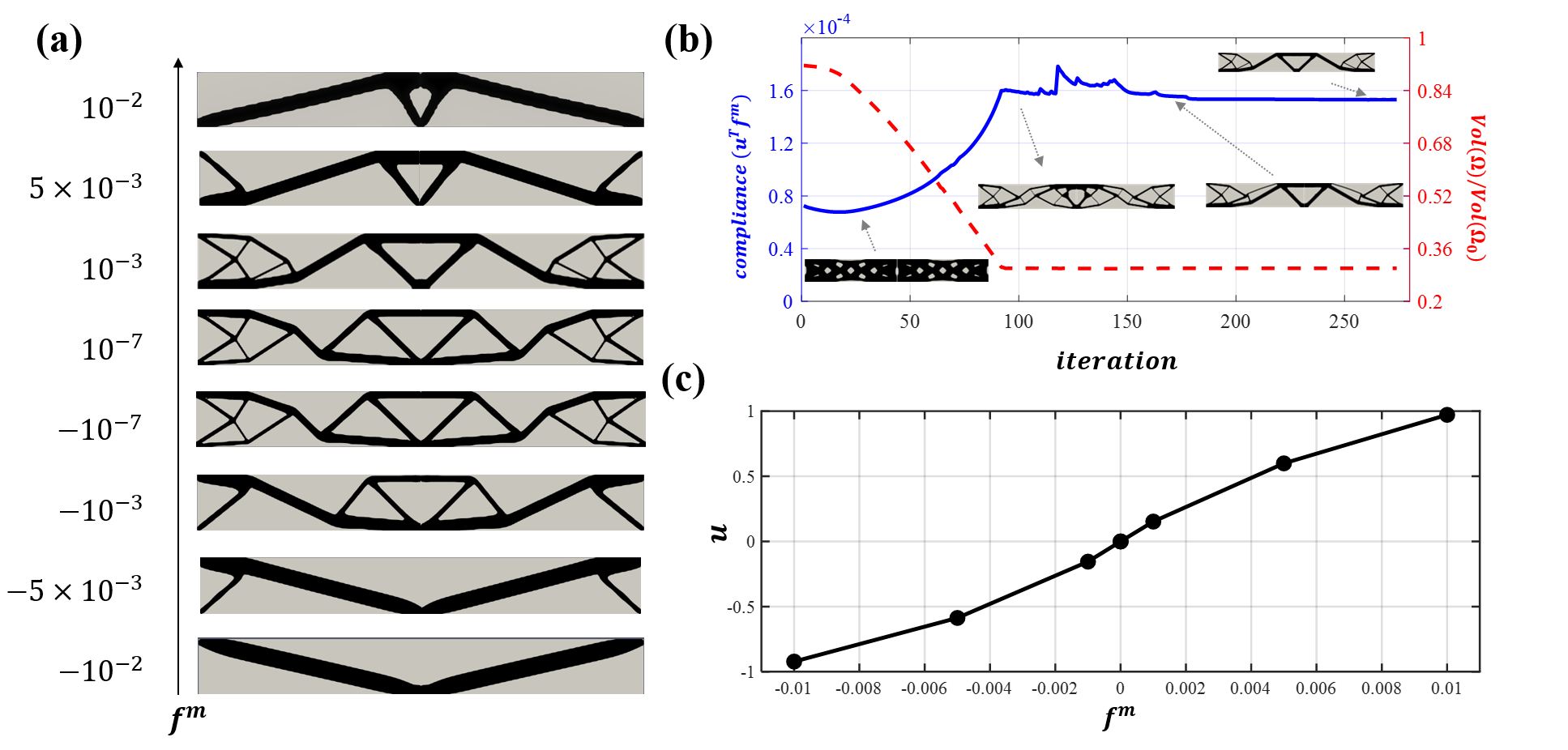}
    \caption{(a) Optimized material layouts for various prescribed external loads $\boldsymbol f^m$; (b) Convergence plot of the case $\boldsymbol f^m = 10^{-2}$, where the intermediate layouts during convergence are plotted as insets; (c) The displacement $u$ at the point of the applied load, calculated at the final equilibrium for various prescribed $\boldsymbol f^m$}
    \label{fig:Res1_mech}
\end{figure} 

The optimized material layouts for the range of the mechanical loads are shown in Fig.~\ref{fig:Res1_mech}. 
For clarity, the whole structural designs are plotted herein by mirroring half of the symmetric design.
Figure \ref{fig:Res1_mech}(a) shows that the optimum solutions gradually change as the mechanical loads $\boldsymbol f^m$ vary. 
When the load is small, i.e., $\|\boldsymbol f^m\| \leq 10^{-7}$, the optimum topology is equivalent to the linear optimum, independent of the load direction. 
This is because both compressive and tensile internal forces induce the same design sensitivities in the small displacement regime. 

As the load magnitude increases, the optimum adapts to a larger deformation by removing slender members in compression,
which in effect prevents the compression-induced buckling. 
When the load is positive (i.e., $\boldsymbol f^m \geq 0$), members at the upper corners are removed as the load increases because the overall structure is pushed up to $+y$ direction.
On the contrary, members at the lower edge are removed in the negative load case and the overall material layout gradually changes to the V-shape as the load becomes more negative. 
Such a removal of the material at the lower edge is a characteristic of the optimum layout of the bi-clamped structure that experiences a negative mechanical loading of a large magnitude \cite{ha2008level, Jung2004}.  

The convergence graph for the case $\boldsymbol f^m = 10^{-3}$ is illustrated in Figure \ref{fig:Res1_mech}(b).
The material is continuously removed until iteration 93 where the volume constraint is satisfied, while the end-compliance increases accordingly.
The materials are rearranged since then, removing many slender members and creating a number of oscillations that are found roughly between iteration 100 to 180. 
The compliance peak, which is shown in iteration 118, is observed during the removal of the compressed members at the center. 
The final convergence is obtained at iteration 272. 

The displacements at the loaded point of the optimum layouts for each load level $\boldsymbol f^m$ are shown in Fig.~\ref{fig:Res1_mech}(c).
The resulting displacements are related monotonically to the given mechanical loading. 
This is expected because the displacement is minimized for a given load level, and the optimum of an intermediate load level should not attain a higher displacement than that of a high load level.

\subsubsection{Thermal and mechanical loading}
\begin{figure}[hbt!]
    \centering
    \includegraphics[width=12cm]{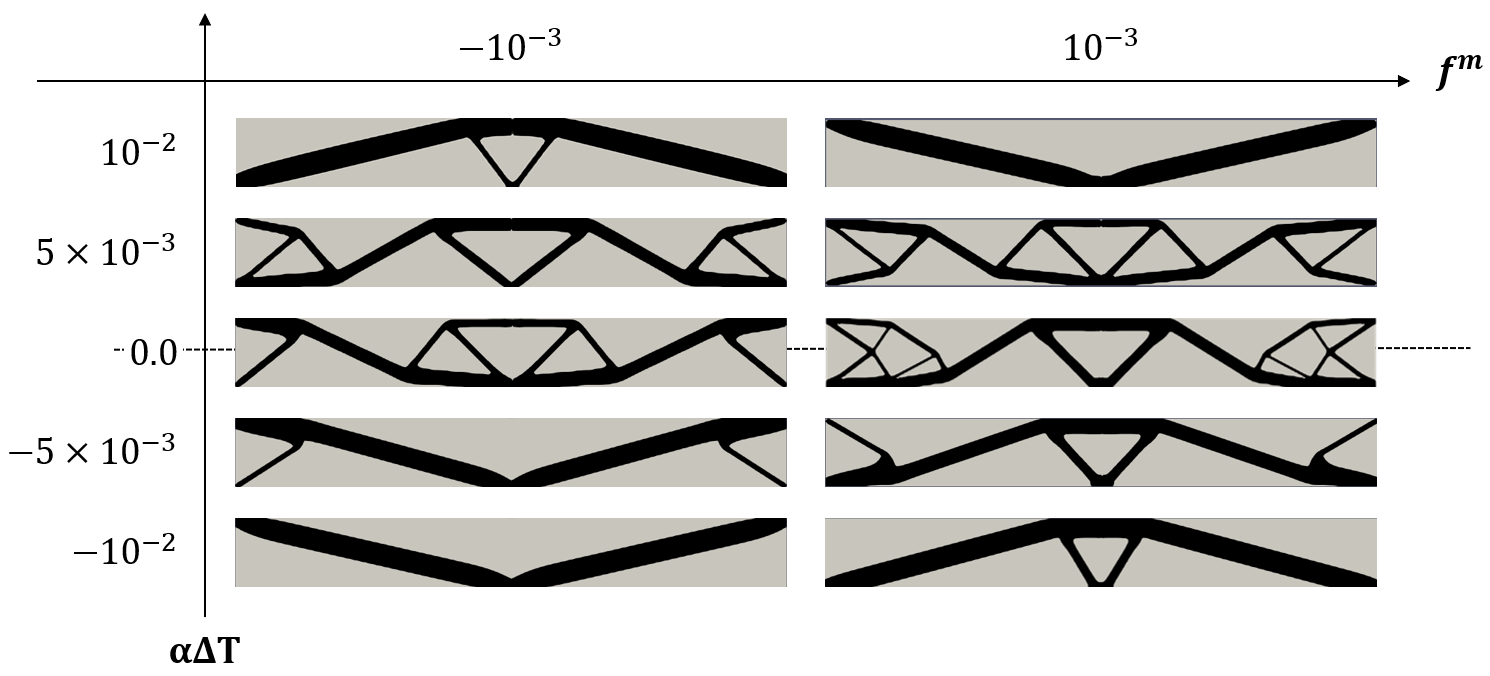}
    \caption{Optimized material layouts considering temperature change and mechanical loads. Mechanical  loads have the same magnitude, but are applied at two opposite directions.}
    \label{fig:Res1_therm}
\end{figure}

We now add a change in temperature, $\Delta T$, which is uniform within the structure. 
The effect is investigated by gradually alternating $\alpha \Delta T$ within the range of $[-10^{-2}, 10^{-2}]$. 
Such a range roughly coincides with the amount of thermal expansion that aerostructures experience in high-temperature operation conditions \cite{Deaton2015}. 
The mechanical load cases are selected to be $[-10^{-3}, 10^{-3}]$, where the given range of thermal loads can balance the mechanical load. 
If the magnitude of the mechanical load $\|\boldsymbol f^m\|$ is higher, i.e.~$10^{-2}$, we found that the material layouts converge to those displayed in Fig. \ref{fig:Res1_mech}, which are mechanically dominant regardless of the temperature change.  
The optimum layouts for the range of temperatures and mechanical loads are illustrated in Fig.~\ref{fig:Res1_therm}. 

Examining the layouts in Fig.~\ref{fig:Res1_therm}, it can be seem that contrary to the results in Fig.~\ref{fig:Res1_mech}, the mechanically compressed members are not always removed. 
One remarkable example is the V-shaped optimized layout for $\boldsymbol f^m = 10^{-3} \text{, when } \alpha \Delta T = 10^{-2}$. 
In addition, the temperature change is shown to affect the design differently depending on the direction of mechanical loading. 
When a negative load is imposed, the layouts changes from V-shape to the inverse-V-shape as the temperature increases. 
The inverse phenomenon is observed when negative temperature is given. 
These changes of optimized designs demonstrate that thermal loading as a design-dependent body load is manipulated via optimization, 
so that the deformation due to temperature change counteracts the mechanical deformation. 

\begin{figure}[hbt!]
    \centering
    \includegraphics[width=12cm]{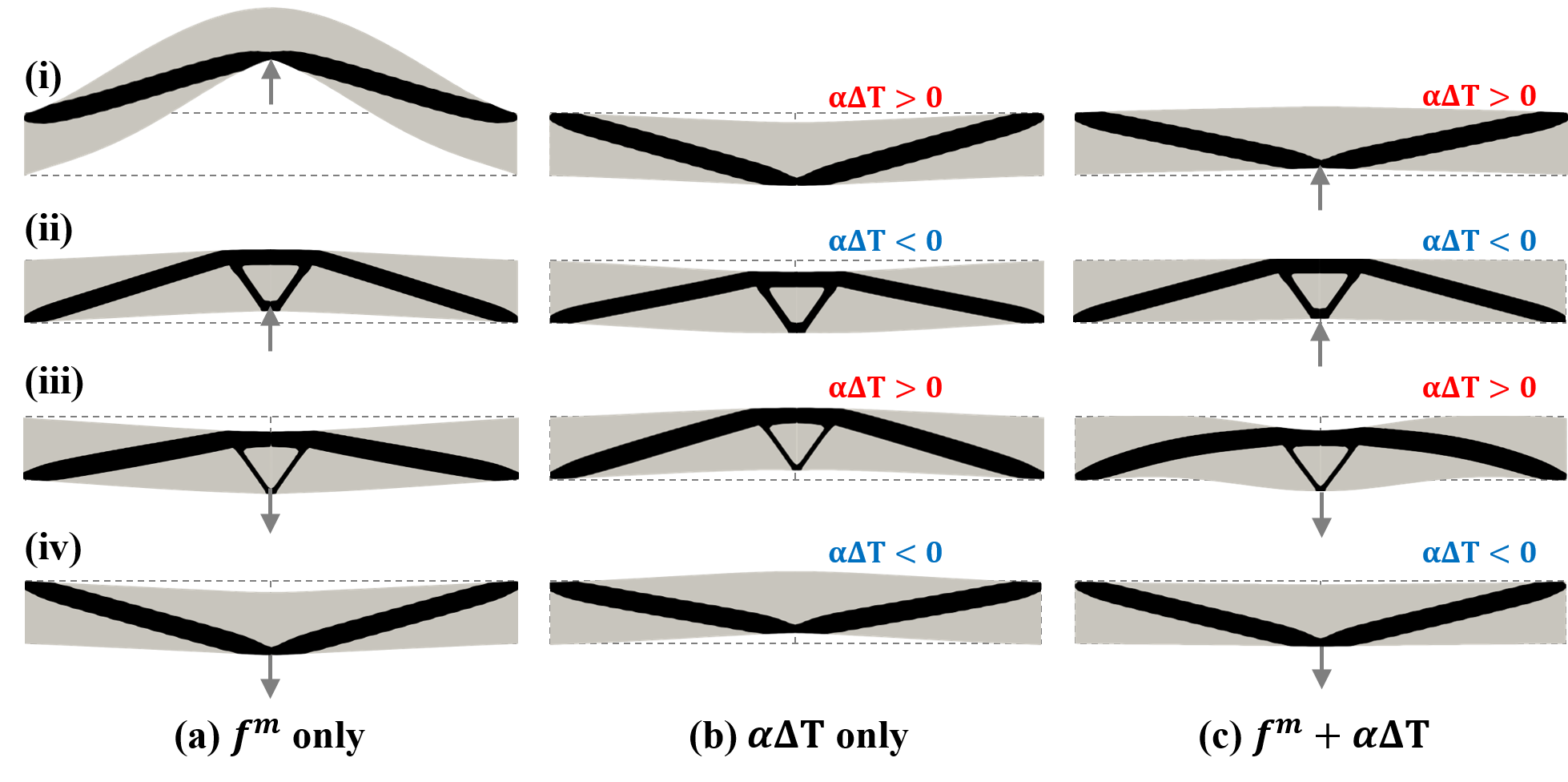}
    \caption{Deformed shapes of four optimum structures found in $\boldsymbol f^m = \pm 10^{-3}$ and $\alpha \Delta T = \pm 10^{-2}$. The signs of mechanical load and temperature change are marked either by arrows and text color. (a) Mechanical load only; (b) Temperature change only; (c) Combined loads.}
    \label{fig:Res1_uth_um}
\end{figure}

To further demonstrate how the temperature change affects the overall structural responses, we study deformations of four designs that are optimized for $\boldsymbol f^m = \pm 10^{-3}$ and $\alpha \Delta T = \pm 10^{-2}$ conditions as shown in Fig.~\ref{fig:Res1_uth_um}. 
The signs of the mechanical loading $\boldsymbol f^m$ and temperature change $\Delta T$ are indicated by a direction arrow and a color of text, respectively. 
For example, the layout found in Fig.~\ref{fig:Res1_uth_um}(i) was obtained from Fig.~\ref{fig:Res1_therm} where the loading condition is $\boldsymbol f^m = 10^{-3} > 0 \text{ and } \alpha \Delta T = 10^{-2} > 0$. 
The deformations of each layout 
are presented, for a mechanical loading only (Fig.~\ref{fig:Res1_uth_um}(a)), a temperature change only (Fig.~\ref{fig:Res1_uth_um}(b)), and a combined thermoelastic loading (Fig.~\ref{fig:Res1_uth_um}(c)). 
These deformations are precisely those that were considered during optimization.  
The dotted line indicates the undeformed design domain. 

By comparing the deformations due to pure mechanical load $\boldsymbol f^m$ to those due to pure temperature change $\Delta T$, it is shown that the layout change is driven towards a result where thermally induced deflection of the tip counteracts the mechanically induced one. 
This reduces the overall deflection hence the objective function. 
This demonstrates how the thermal effect can be exploited and significantly manipulated in order to achieve a certain design goal. 
At an elevated temperature, the tip node of the V-shape and the inverse-V-shape each displaces to $+y$ direction (Fig.~\ref{fig:Res1_uth_um}(b)(i)), and $-y$ direction (Fig.~\ref{fig:Res1_uth_um}(b)(iii)). 
The opposite phenomenon is observed when negative temperature change is imposed (Fig.~\ref{fig:Res1_uth_um}(b)(ii) and Fig.~\ref{fig:Res1_uth_um}(b)(iv)). 
This in effect, counteracts the effect of mechanical loads on the structures (Fig.~\ref{fig:Res1_uth_um}(a)) hence minimizes overall displacements (Fig.~\ref{fig:Res1_uth_um}(c)). 
It is remarkable that even a mechanically induced snap-through behavior of the V-shape (Fig.~\ref{fig:Res1_uth_um}(a)(i)) is suppressed by the temperature effect. 
This demonstrates that topology optimization considering thermoelastic nonlinearity opens up many possibilities for controlling and manipulating large deformations by inducing temperature changes.

\begin{figure}[hbt!]
    \centering
    \includegraphics[width=12cm]{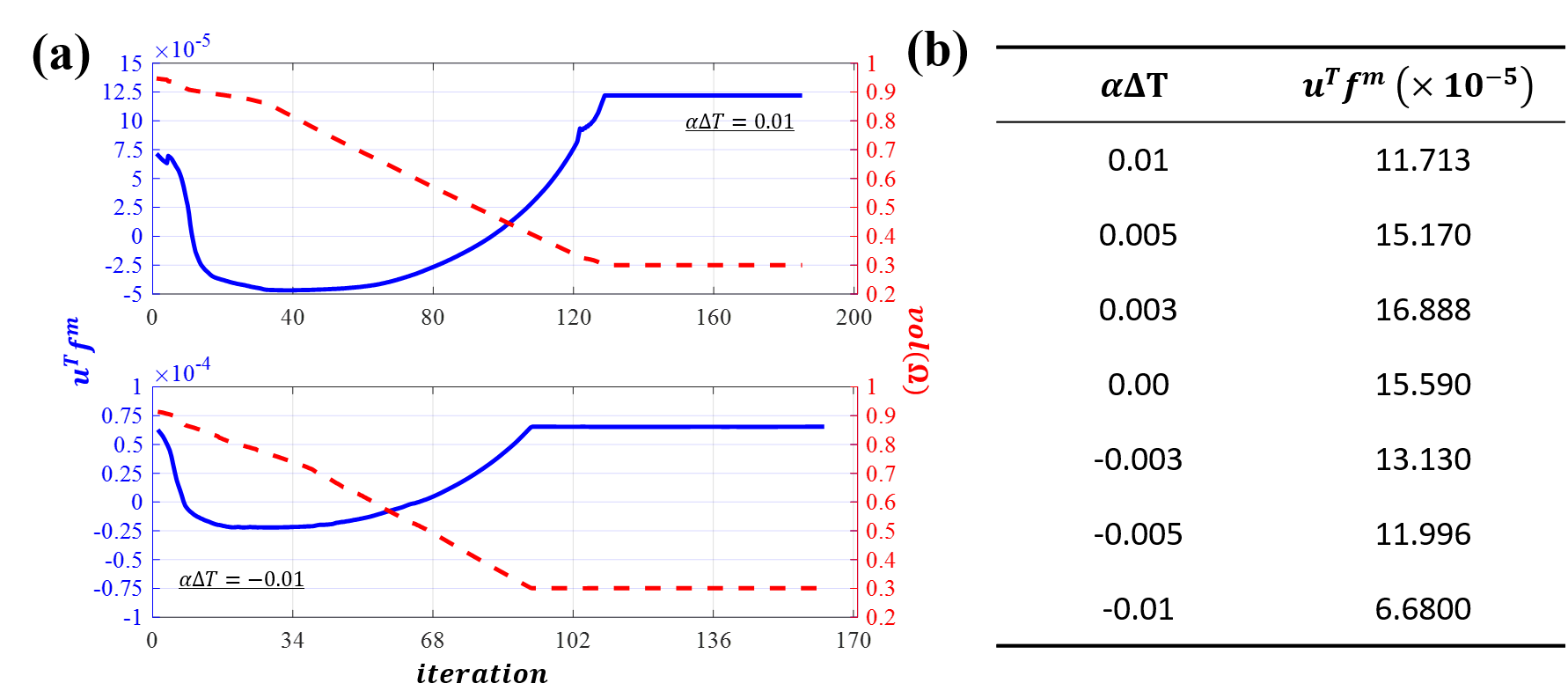}
    \caption{(a) Convergence graphs of the case $\boldsymbol f^m = 10^{-3}$ with negative and positive $\alpha \Delta T$; (b) Optimum compliance values obtained in different temperature conditions.}
    \label{fig:Res1_therm_conv}
\end{figure} 

The convergence graphs and the end objective values for $\boldsymbol f^m = 10^{-3}$ and $\alpha \Delta T = \pm 0.01$ are shown in Fig.~\ref{fig:Res1_therm_conv} to shed more light on the way temperature changes affect optimum results.
When a nonzero temperature change is prescribed, the convergence graphs (Fig.~\ref{fig:Res1_therm_conv}(a)) demonstrate a wide region where the reduction of the material does not necessarily generate a higher displacement. 
For example, the increase of the end compliance is not initiated until the material reduced to approximately 70\% of total domain achieved at iteration 56 (Fig. \ref{fig:Res1_therm_conv}(a), $\alpha \Delta T = 0.01$) and 43 (Fig. \ref{fig:Res1_therm_conv}(a), $\alpha \Delta T = -0.01$). 
Such a negative contribution of the material reduction is due to thermally-driven displacements found in an early stage of the iterations. 
Such a suppressed deflection when compared with the mechanical-only case, is a possible explanation for attaining the same design regardless of the assumed symmetry. 
If an instability is dominant in structural behavior, e.g., the bi-clamped beam with an arc, the effect of the assumed symmetry is expected to be have a greater influence on the optimum design.

\begin{figure}[hbt!]
    \centering
    \includegraphics[width=12cm]{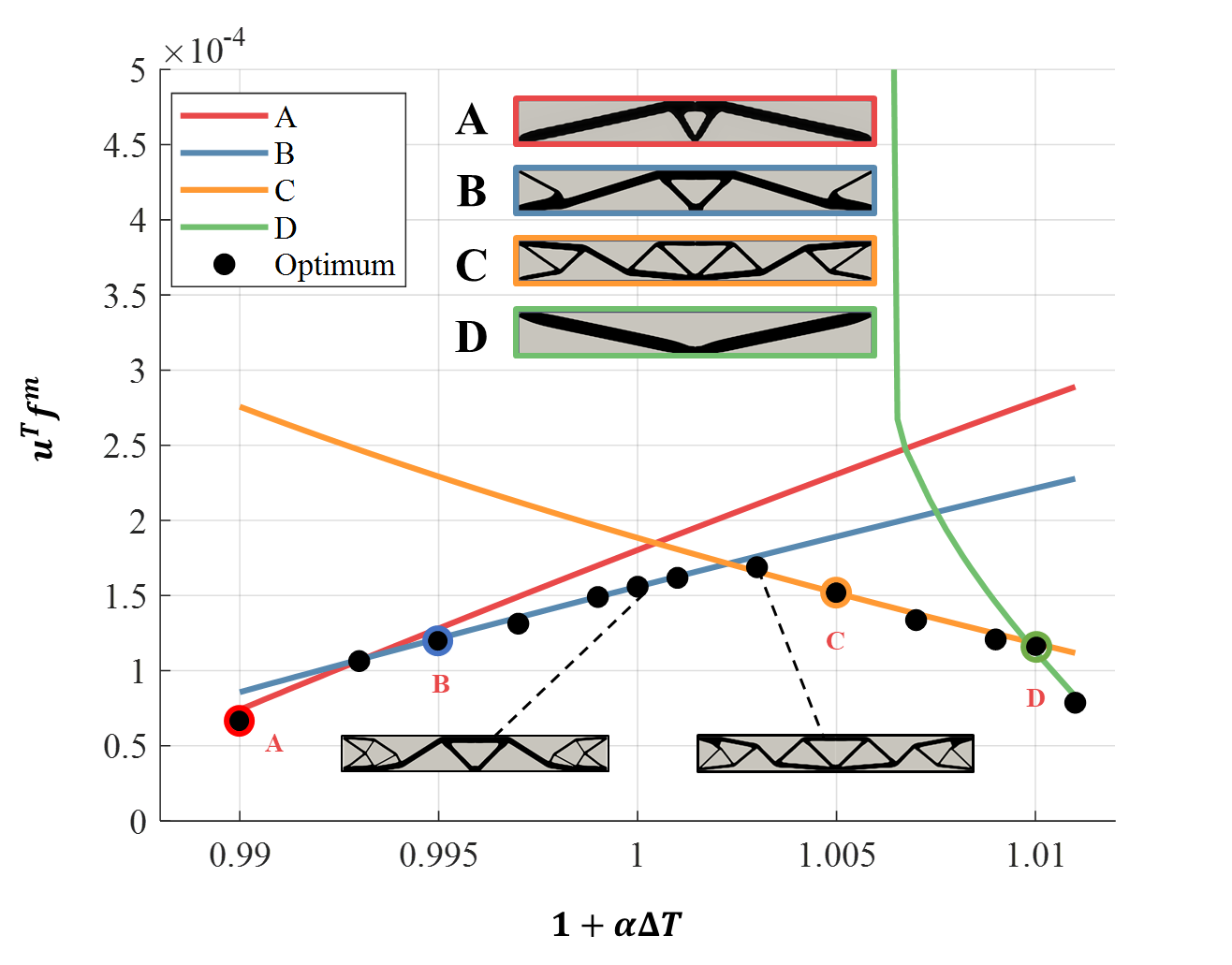}
    \caption{The thermomechanical responses of different material layouts analyzed by finite element analysis for varying temperature condition $\alpha \Delta T$. 
    The behaviors of the representative layouts (A-D) are color-coded. 
    The optimized results of these layouts are marked by the same colored dots, and finite element analysis results are plotted as a line, where the temperature effect $\alpha \Delta T$ is incremented by $10^{-3}$.   
    The black dots indicate the optimized value for additional temperature quantities other than those that correspond to the given layouts. 
    \color{blue} 
    }
    \label{fig:Res1_envelope}
\end{figure} 

The effect of thermal loads on topological design under large deformations is further examined by a parametric study shown in Fig.~\ref{fig:Res1_envelope}. 
The analysis is based on nonlinear thermoelastic finite element analysis of selected optimum layouts. 
Four representative optimum layouts are selected among the optimum layouts for the $\boldsymbol f^m = 10^{-3}$ case:  
the layouts \textit{A, B, C, D} correspond to $\alpha \Delta T = -0.01, -0.005, 0.005, 0.01$ cases. 
Each of these layouts has a distinctive topology to which the nonlinear 
finite element analysis is applied.
For a consistent comparison the mechanical load is fixed at $\boldsymbol f^m = 10^{-3}$ and the temperature condition is varied $\alpha\Delta T = [-0.01,0.011]$ by increments of $10^{-3}$.
The compliance values of these layouts are calculated based on the structural response $\boldsymbol u$ for given $\boldsymbol f^m$ and $\alpha \Delta T$.
The values are plotted with respect to $\alpha \Delta T$ as continuous lines, which are color-coded with the corresponding layouts shown in the inset. 
As the compliance profile with respect to the temperature changes depends upon the curvature direction of the layouts, which agrees with the thermally induced displacement described by Fig.~\ref{fig:Res1_uth_um}, a convex design envelope is created.
The optimum compliances reported in Fig.~\ref{fig:Res1_therm_conv}(b) are marked as black dots. 
This figure demonstrates that 
for any level of $\alpha \Delta T$, the design optimized for a specific $\alpha \Delta T$ outperforms the other designs, indicating that the solutions obtained by the proposed method are optimum.

Fig.~\ref{fig:Res1_envelope} also shows the significance of the design envelope in the context of the nonlinear thermoelastic problem.
The optimum compliance values that result from optimizing at 13 different level of $\alpha \Delta T$ are shown to be highly correlated with the envelope created by the intersecting lines.
The deviation between the optimal value with the envelope is originated from a gradual transition between layouts, e.g., $\Delta T =0$ case between \textit{B} and \textit{C}, but the maximum deviation is found to be less than 10\%. 
Notably, such a correlation holds even when there is a drastic switch between the layouts near $\alpha \Delta T = 0.003$, where the analyzed results of layouts \textit{B} and \textit{C} intersect. 
At the intersection, predominantly \textit{C}-like material layouts are obtained (see the inset of Fig.~\ref{fig:Res1_envelope}).
It is clear that the nonlinearity is well captured in the envelope even to the extent of the instability, as shown in the finite element results of layout \textit{D}.
This structure goes through snap-through around $\alpha \Delta T = 0.007$, 
which is suppressed by a higher temperature, $\alpha \Delta T > 0.007$.

\subsubsection{Effect of nonuniform temperature}

\begin{figure}[hbt!]
    \centering
    \includegraphics[width=12cm]{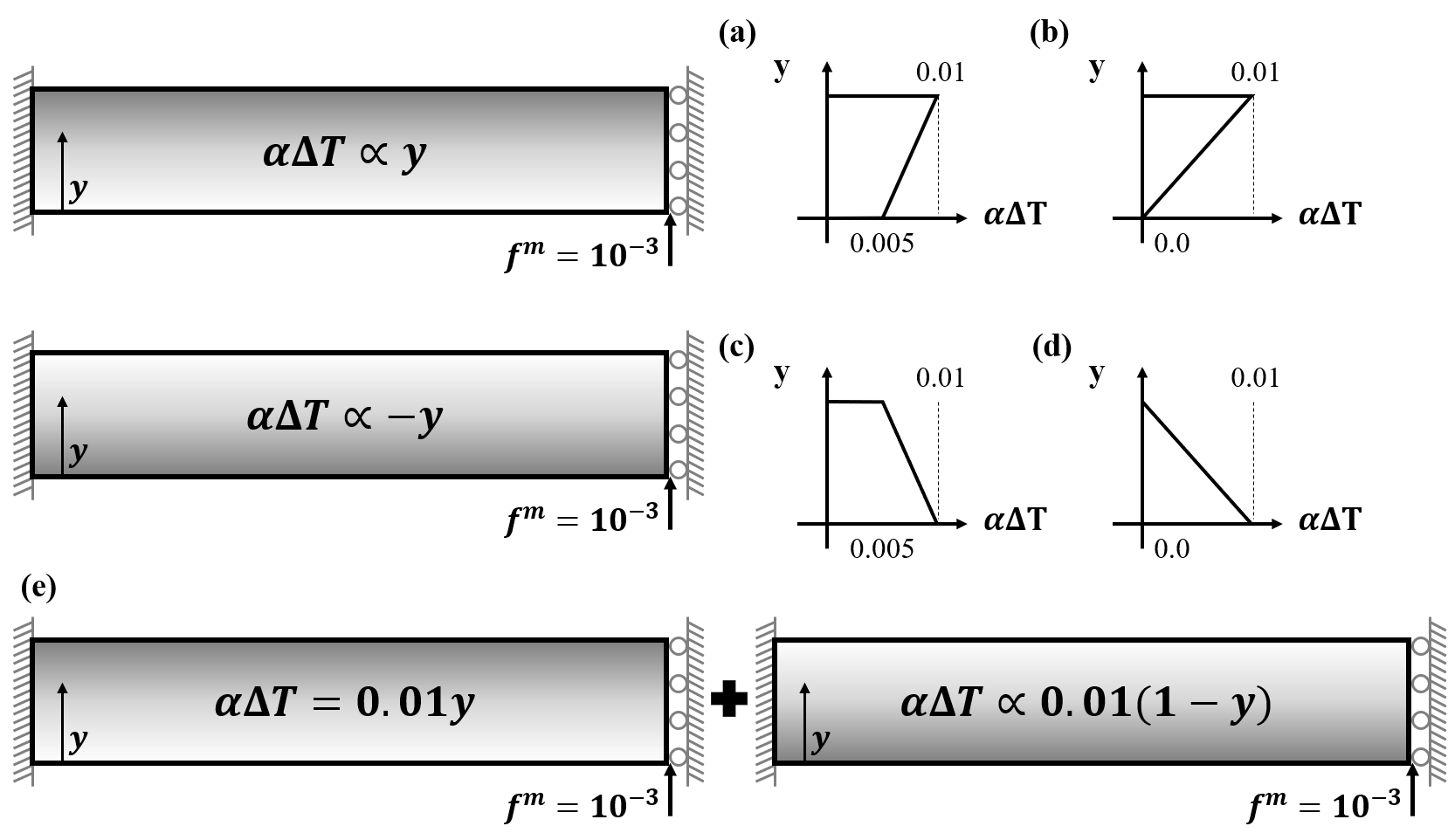}
    \caption{A set of temperature distributions, where the numbers indicate an effective thermal expansion ratio ($\alpha\Delta T$). (a-d) single temperature distribution case; (e) weighted sum of two temperature distribution cases, which is analogous to multiple load cases.
    }
    \label{fig:R1_ex1_config}
\end{figure}

The effect of non-uniform temperature change $\Delta T$ is investigated by spatially varying the strain $\alpha \Delta T$.
The mechanical load $\boldsymbol{f}^m$ acts in the $+y$ direction with a magnitude of ${10}^{-3}$. 
\color{black}
The thermal strain $\alpha\Delta T$ is assumed to vary in the $y$-direction so that the induced thermal expansion is expected to create a bending moment, which affects the optimization.
\color{black}
To avoid any inconsistency between the polynomial order of thermal and mechanical strain fields, $\alpha \Delta T$ is assumed to be piecewise constant in each element found in the finite element grid and is evaluated at the element centroid. 
The distribution of $\alpha \Delta T$ is assumed to be fixed throughout optimization. 
Five different temperature gradients are examined as shown in Fig.~\ref{fig:R1_ex1_config}. 

\begin{figure}[hbt!]
    \centering
    \includegraphics[width=12cm]{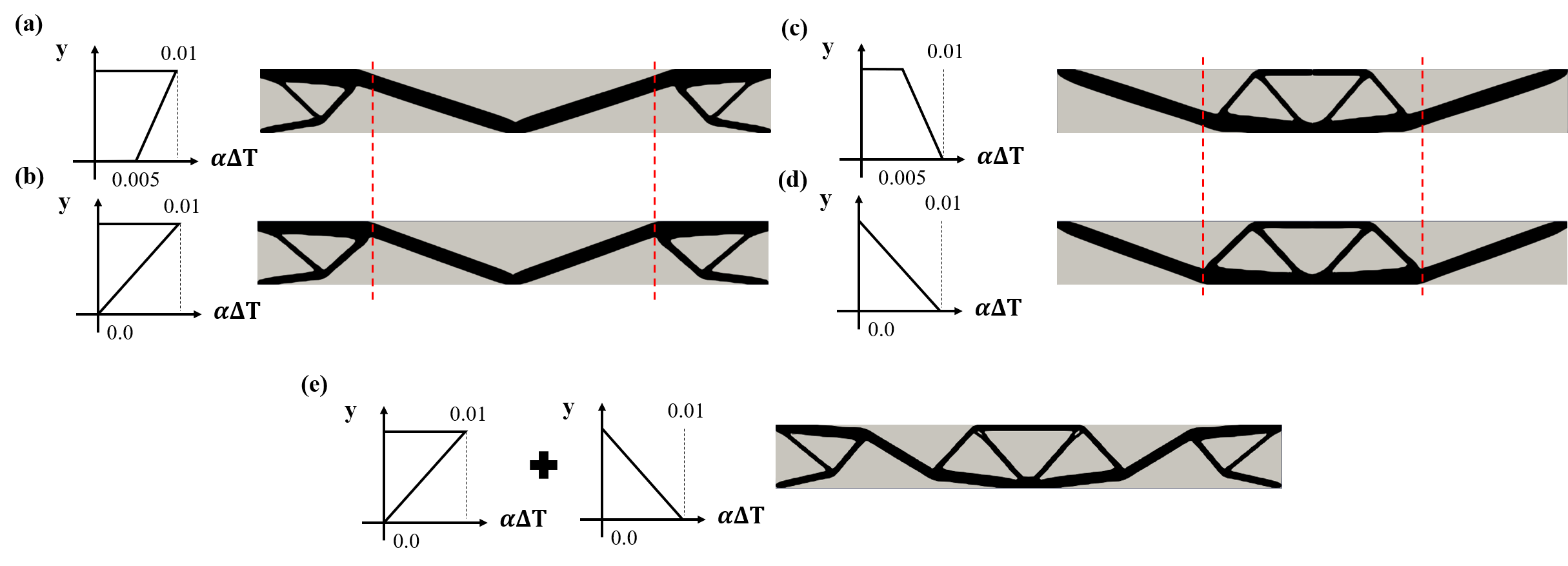}
    \caption{Optimized material layouts considering nonuniform distribution of temperature change $\alpha\Delta T$, which correspond to Fig.~\ref{fig:R1_ex1_config}.
    The red dotted lines are drawn to highlight the influence of the temperature gradients on the optimized structures. 
    }
    \label{fig:R1_ex1_results}
\end{figure}

Both increasing (Fig.~\ref{fig:R1_ex1_config}(a, b)) or decreasing (Fig.~\ref{fig:R1_ex1_config}(c, d)) $\alpha \Delta T$ with respect to the $y$-coordinates are examined herein, as well as the effect of \color{black}their linear gradient.
To examine the effect of considering multiple thermal load cases, two temperature distributions with different gradient signs are considered as separate load cases as shown in Fig.~\ref{fig:R1_ex1_config}(e) and
\color{black}
the optimum design is expected to accommodate different thermal load scenarios.
The optimum layouts for the various thermal distributions are illustrated in Fig.~\ref{fig:R1_ex1_results}. 

\color{black}
An examination of the optimum layouts in Fig.~\ref{fig:R1_ex1_results} with the uniform $\alpha\Delta T$ conditions---in particular, the layouts in the top two rows, right hand side of Figure~\ref{fig:Res1_therm}---illustrates how the non-uniform temperature affects the optimum layouts.
All of the designs found in Fig.~\ref{fig:R1_ex1_results} (a) through (d) predominantly consist of diagonal beam-like members, which are obtained as the optimum when $\alpha \Delta T=0.01$ is imposed uniformly onto the design domain (see Fig.~\ref{fig:Res1_therm}). 
As $\alpha\Delta T$ becomes uniform (i.e., (b) to (a), or (d) to (c)), it is clearly seen that the optimum structure gradually converges to the V-shaped structure. 
\color{black}
In addition, the locations of the diagonal members within the structure are shown to be a function of the sign of the gradient. With a positive gradient (Fig.~\ref{fig:R1_ex1_results}(a, b)), the loaded region is directly connected to the thin diagonal members, which makes the structure prone to buckle when compared with designs optimized for the negative gradients (Fig.~\ref{fig:R1_ex1_results}(c, d)). 
\color{black} 
As the direction of the curvature in Fig.~\ref{fig:R1_ex1_results}(a, b) is opposite to the direction in Fig.~\ref{fig:R1_ex1_results}(c, d), optimization yields different buckling-resistant topologies: 
a V-shape in the former case, Fig.~\ref{fig:R1_ex1_results}(a ,b), versus with the trapezoidal stiffener in the center in the latter case, Fig.~\ref{fig:R1_ex1_results}(c, d).
These results indicate the direction of the gradient has an effect, not to mention the total amount of thermal loads (i.e., $\alpha \Delta T$).
Another interesting observation is that these two distinct topologies, the V-shaped structure versus the layout with the central trapezoidal stiffner, borrow the topological features from the design for uniform $\alpha \Delta T=0.005$, which is also the average change of the results in Fig.~\ref{fig:R1_ex1_results}(b) and (d).
Interestingly, the multiple thermal load case in Fig.~\ref{fig:R1_ex1_results}(e) is topologically identical to the result for the same uniform temperature in Fig.~\ref{fig:Res1_therm}.
This is not an obvious result; when only mechanical multiple load cases are considered, the resulting solution is usually different from when they are considered as a single load case. 
We note that in Fig.~\ref{fig:R1_ex1_results}(e), the mechanical load cases are the same but the thermal load cases are different ---this might explain the convergence towards the same layout as for the uniform temperature case.

\color{black}

\subsection{Square design domain with a central load}
\begin{figure}[hbt!]
    \centering
    \includegraphics[width=12cm]{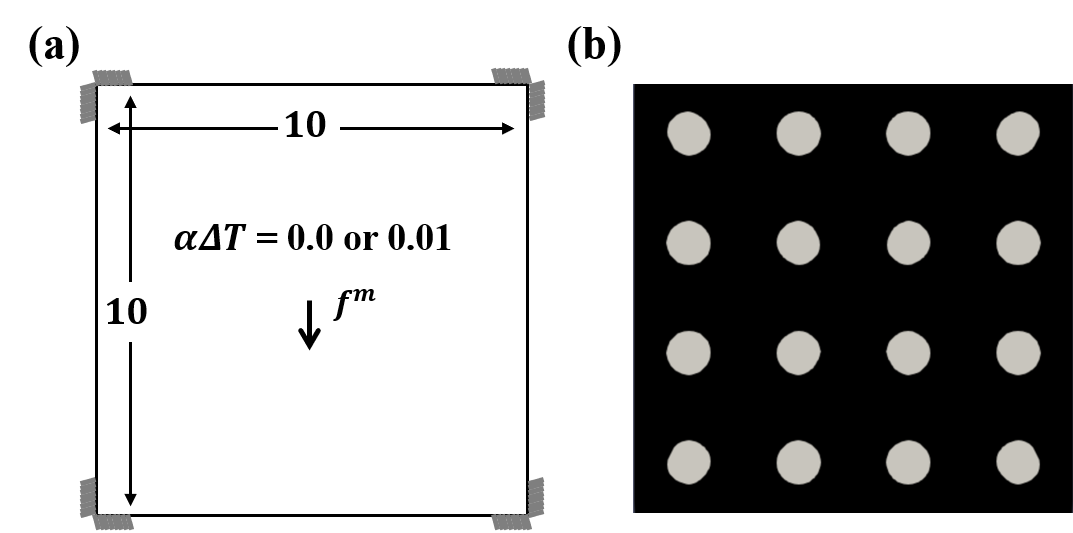}
    \caption{Design configuration of the square design domain with a central load. (a) dimensions and boundary conditions of the design problem; (b) initial material distribution, where the void regions are colored in gray.
    }
    \label{fig:R1_ex2_config}
\end{figure}

The square domain is considered as shown in Fig.~\ref{fig:R1_ex2_config}(a), to further demonstrate the counteraction between thermal and mechanical effects. 
The thermal expansion ratio $\alpha\Delta T$ is assumed to be uniform in this case. 
A unit mechanical force $\boldsymbol{f}^m$ is applied at the center of the square, and the final volume is set to 30\% of the initial material volume. 
The clamped boundary conditions are imposed in the regions within a distance of 0.5 from the corners. 
The design domain is discretized by 10,000 quadrilateral elements. 
The initial material layout is shown in Fig~\ref{fig:R1_ex2_config}(b), where the void areas are shown in gray and the material region in black.

\begin{figure}[hbt!]
    \centering
    \includegraphics[width=12cm]{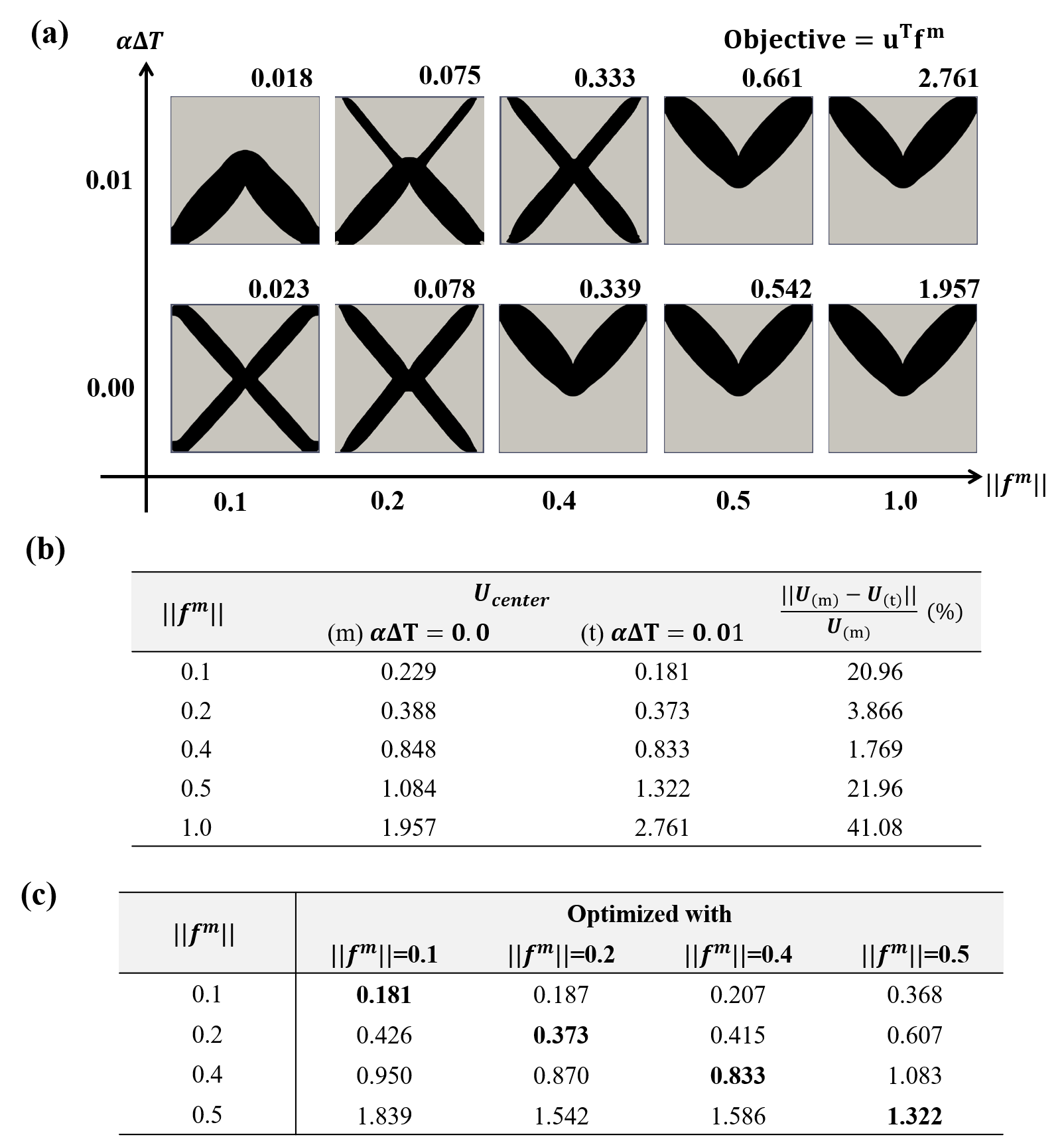}
    \caption{
    Optimum results for the variation of temperature changes $\alpha\Delta T$ and mechanical loads $\boldsymbol{f}^m$. (a) optimized layouts for the thermoelastic loads and the objective values; (b) central displacements and the normalized differences between the coupled case (t) and pure mechanical case (m); \color{black}(c) validation of the optimality of the layouts optimized with different $\boldsymbol{f^m}$, by comparing deflections induced by different $\boldsymbol{f^m}$. The layout with the smallest deflection is marked by bold as it is the optimum, showing that each layout is indeed the best performer under its own assumed conditions.
    }
    \label{fig:R1_ex2_result}
\end{figure}

In the present example, both mechanical and thermal effects are investigated by imposing two thermal expansions, i.e., $\alpha\Delta T=$ 0.0 or 0.01, while changing the magnitude of the mechanical loads $\boldsymbol{f}^m$ from 0.1 to 1.0. 
Figure~\ref{fig:R1_ex2_result}(a) presents the optimized material layouts for five thermoelastic loads, along with the end-compliances $\boldsymbol{u}^T\boldsymbol{f}^m$ evaluated at the last iteration. 
As noted earlier, the present objective function is employed to optimize the structure that minimally deflects under the given thermoelastic loads, hence the optimizer is expected to design a structure with the least possible deflection. 
The vertical displacements at the loaded node are computed and listed in Fig.~\ref{fig:R1_ex2_result}(b).

Both mechanical and thermal loads are found to significantly affect the optimized solutions. 
In the case of a mechanical loading problem with $\alpha\Delta T$ = 0.0, the optimum layout changes from a X-shape to a V-shape as the mechanical load increases. 
The X-shape which induces a deformation of roughly 2\% of the dimension, agrees well with a linear compliance minimization problem found in Ref.~\cite{Takalloozadeh2017}. 
The V-shape, on the other hand, is obtained when the deformation roughly exceeds 8\%. 
In this nonlinear regime, the bottom members are removed since they are prone to buckle as the load increases. These results are consistent with the numerical solutions found in Fig.~\ref{fig:Res1_uth_um}. 

When $\alpha\Delta T=0.01$ is imposed with the same range of the mechanical loads, \color{black}an inverse V-shape is obtained as the optima when the isolated mechanical load induces a small deformation, i.e., deflection being smaller than 5\% of the dimension. 
When $\|\boldsymbol{f}^m\|$ is further increased to the degree where the V-shape is optimum when the mechanical-only case, the bottom members are reduced and hence X-shape is recovered to be an optimum. 
\color{black}
The bottom members, of which thermal expansion deflects the structure in the $+y$ direction (see Fig.~\ref{fig:Res1_therm_conv} (iii)), effectively mitigate the mechanical deflection (i.e., $U_{\left(m\right)} > U_{\left(t\right)}$). 
The degree of mitigation is gradually diminished as the thickness of the bottom members are reduced, as the reduction ratio (i.e. $\|\boldsymbol{U}_{\left(m\right)}-\boldsymbol{U}_{\left(t\right)}\|/\boldsymbol{U}_{\left(m\right)})$ changes from 20.96 \% to 1.769 \% as $\|\boldsymbol{f}^m\|$ changes from 0.1 to 0.4.
When $\|\boldsymbol{f}^m\| \leq 0.5$, the V-shape is found to as the optimum as it avoids snapping behavior induced by the increased $\|\boldsymbol{f}^m\|$.
In this range, the design is dominantly affected by mechanical loading, which is in the $-y$ direction.
Therein, $\boldsymbol{U}_{\left(m\right)}<\boldsymbol{U}_{\left(t\right)}$ because the thermal deformation also deflects the layout in the $-y$ direction. 

\color{black} 
To demonstrate the optimality of the results, the deflections of the layouts at various $\boldsymbol{f^m}$ are shown in Fig.~\ref{fig:R1_ex2_result}(c). 
Four different layouts that are optimized with different $\|\boldsymbol{f^m}\| \in [0.1, 0,2, 0,4,0.5]$ are analyzed by nonlinear finite element analysis with the specified $\|\boldsymbol{f^m}\|$ along with the elevated temperature condition (i.e., $\alpha \Delta T=0.01$). 
The deflections at the center node are tabulated in Fig.~\ref{fig:R1_ex2_result}(c), where 
the layouts with the smallest deflections are highlighted and they are the optima among the layouts considered herein. 

When $\|f^m\| \leq 0.2$, the inverse-V shape or its variation with slender diagonal members in the upper half are optimal as the members found in their lower half generate a structural deflection in the direction opposite to the mechanical loads. 
Such thermally-driven counteraction to the mechanical load is also observed in the inverse-V shape found in bi-clamped beam example, as shown in Fig.~\ref{fig:Res1_uth_um}. 
When the mechanical load is further increased (i.e., $\|f^m\| \geq 0.4$), however, the members found in the lower half (i.e. inverse-V shape) is not beneficial as they are not effective load-carrying structures when compared with the V-shaped structure.
As a result, the layouts that have less or no members at the lower half of the domain are found to be the least deforming structure for the corresponding load hence are the optima.  
\color{black}

\subsection{Uniaxial beam under compressive load}

\begin{figure}[hbt!]
    \centering 
    \includegraphics[width=8cm]{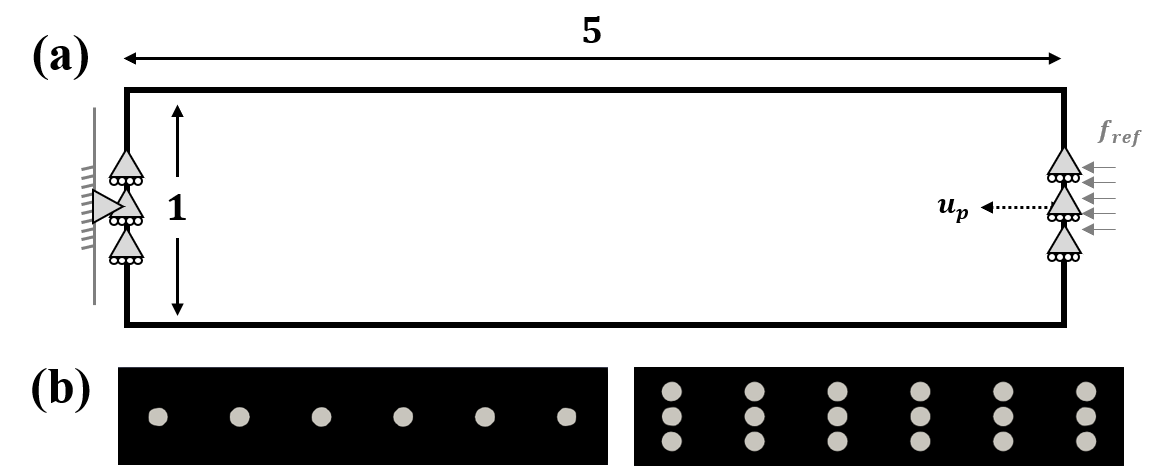}
    \caption{Design configuration of the uniaxial beam. (a) Dimensions and boundary conditions of the problem; (b) Initial material distributions, where the void material is marked by gray.}
    \label{fig:Res2_config}
\end{figure}

We now examine thermoelastic structural design for the case of an inherent instability. 
A simply supported beam under uniaxial compression is presented in Fig.~\ref{fig:Res2_config}.
A length of 0.3 of each edge is pinned, 
analogous to the Euler beam buckling.
The displacement-control scheme is used to find the equilibrium, because the intermediate solutions with reduced material volume are expected to become unstable. 

The problem (\ref{eqn:compliance_dispCtrl}) is solved with material volume limited up to 50\% of the total design domain. 
The controlled displacement $u_p$ is specified at the center node on the right boundary, while the uniform reference force $\hat{\boldsymbol f^m}$ is distributed over the length of 0.3. 
The support and loading regions are specified as non-designable. 
To prevent a sharp singularity at the structural instability, a load imperfection is imposed by uniformly translating the reference force in the $+y$ direction by 0.025. 
The domain is discretized by a 32000 square quadrilateral finite element mesh. 
To simulate the buckling without loss of generality, symmetry and a specific mode of buckling are not assumed during the analysis, whereas double symmetry is imposed on the design domain. 
As shown in Fig.~\ref{fig:Res2_config}(b), 6 or 18 circular material voids are initially imposed along the centerline. 

\subsubsection{Mechanical loading only}

In order to obtain the reference results for the thermoelastic optimization case, we first solve the mechanical design problems without temperature change.
In topology optimization under linear elasticity, the minimum compliance structure under compressive uniaxial loading is a simple straight beam whose thickness is determined only by the volume constraint \cite{dunning2016level}. 
With nonlinearity, optimization captures the buckling instability, where the load carrying capacity is greatly reduced.

\begin{figure}[hbt!]
    \centering
    \includegraphics[width=12cm]{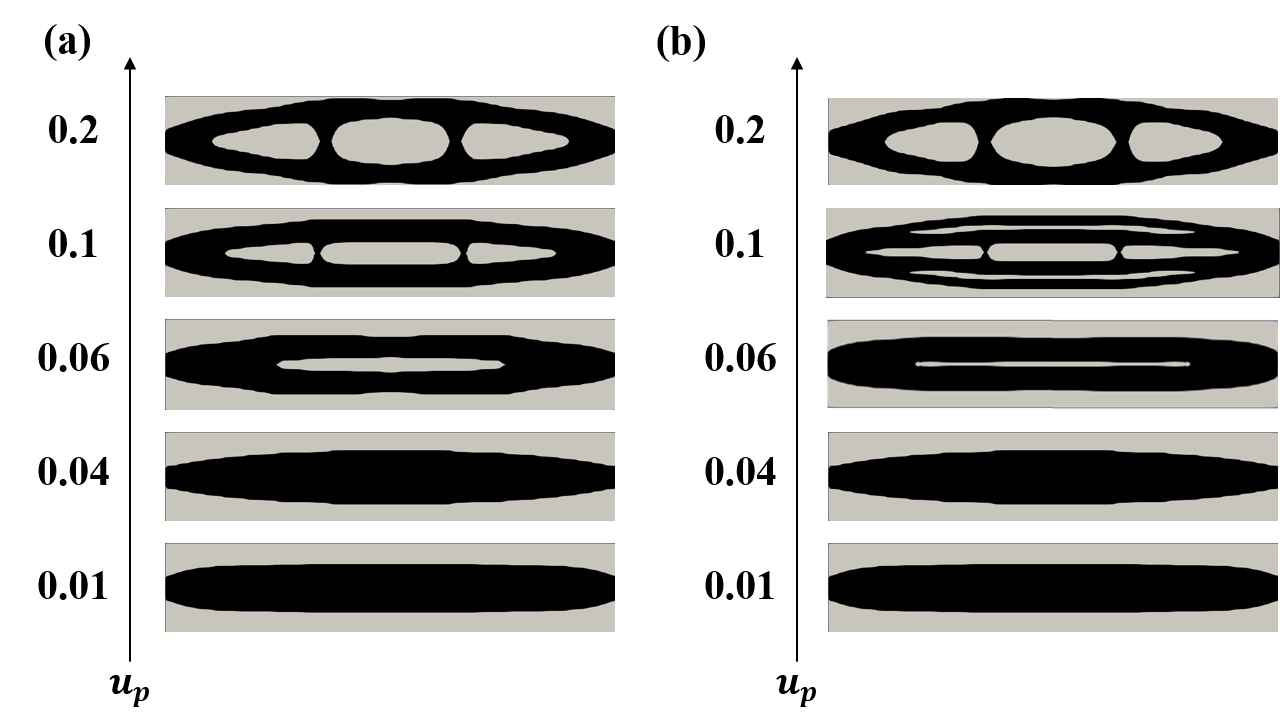}
    \caption{Optimum layouts for various values of $u_p$. (a) Optimum solutions with 6-hole initial configuration; (b) Optimum solutions with 18-hole initial configuration.} 
    \label{fig:Res2_mech}
\end{figure} 

Figure \ref{fig:Res2_mech} shows the optimum structures for a range of applied displacement $u_p$ starting with two different initial configurations of Fig. \ref{fig:Res2_config}(b). 
The final topologies are found to be similar, with objective function differences less then 1.5\% in all cases, showing that the effects of the initial void configuration is not significant for this investigation.

The changes between these layouts effectively demonstrate how the optimization scheme produces the designs that are optimum for the increasing levels of displacement. In case of the smallest displacement $u_p = 0.01$, the optimum layout is a solid beam, which has maximum compressive stiffness for the given volume constraint. 
This layout agrees well with the linear solution in \cite{dunning2016level}. 
As shown in the layouts obtained at $u_p = 0.04$, increasing $u_p$ leads to materials to be distributed further away from the neutral axis, hence the buckling resistance of the solid beam is enhanced. 
Such solid topological layout is retained until $u_p \leq 0.06$, which corresponds to 1.2\% compressive strain. 
At that point, a slit is created at the center of the structure, further enhancing buckling rigidity for the given volume constraint. 
For $u_p \geq 0.1$, additional struts are added such that the region with higher cross-section moment of inertia is enlarged.  

\begin{figure}[hbt!]
    \centering
    \includegraphics[width=12cm]{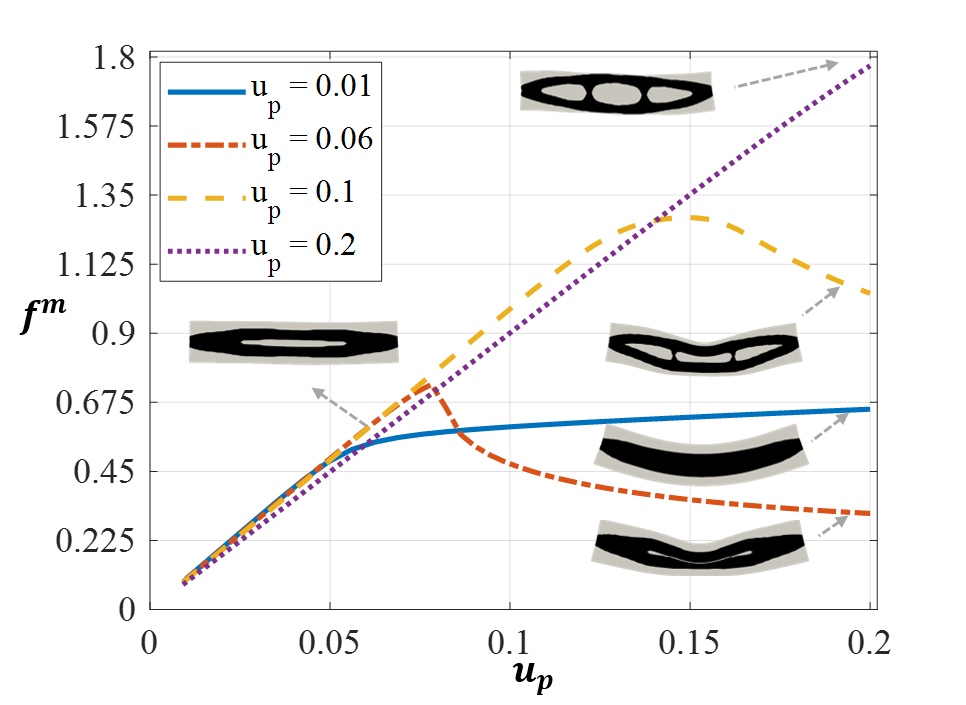}
    \caption{Force-displacement responses of four optimum designs, showing their \color{black} superior response at the displacement level for which they were optimized, \color{black} and their post-buckling response at a larger displacement.}
    \label{fig:Res2_discuss}
\end{figure} 
We compare the mechanical responses of the optimum layouts under the largest prescribed displacement $u_p = 0.2$
as shown in Fig.~\ref{fig:Res2_discuss}. 
Four layouts (obtained with $u_p = 0.01, 0.06, 0.1, 0.2$) in Fig.~\ref{fig:Res2_mech}(a) are chosen for this investigation. 
The resulting force-displacement graphs that are obtained from the finite element analyses are plotted. 
The results demonstrate that the solution optimized for the higher $u_p$ avoids instabilities, while having a smaller load-carrying capacity in low $u_p$.  
The deformed shapes are shown in the inset of Fig.~\ref{fig:Res2_mech}(b).
Up to $u_p \leq 0.06$, 
the responses of the representative layouts are all linear.
Creating a slit when $u_p \geq 0.06$ delays the onset of buckling.
Ultimately, the layout obtained for $u_p = 0.2$ is shown to be capable of sustaining 70\% more load than the solid beam structure at $u_p = 0.2$.
According to these results, it can be therefore concluded that 
the proposed optimization method produces the slit and struts, which are beneficial in maximizing buckling capacity for the given material volume. 
The study also helps to understand why slits and struts are found in optimum structures when mechanical buckling load is considered through a linear buckling constraint \cite{Deng2017}, or randomly imposed geometric imperfections \cite{Jansen2014}. 
When the present scheme is incorporated within the multiscale optimization \cite{Sivapuram2016}, hierarchical lattice-like structures are expected as optima in the small-scale regime \cite{thomsen2018buckling}. 
For example, beams under compressive loads are divided into a set of struts and voids in order to further present the local buckling therein.


\subsubsection{Thermal and mechanical loading}

\begin{figure}[hbt!]
    \centering
    \includegraphics[width=12cm]{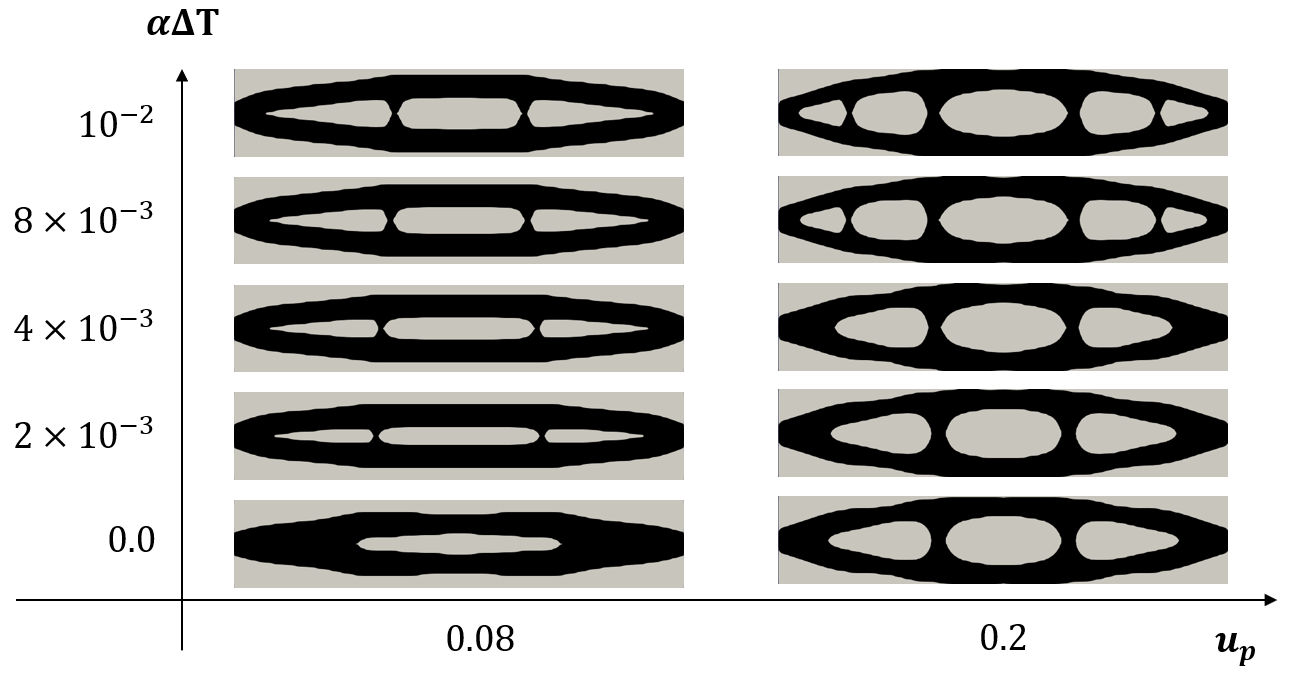}
    \caption{Optimum material layouts under large deformations and temperature change. The salient topology shift observed as $\alpha \Delta T$ changes. 
    }
    \label{fig:Res2_therm}
\end{figure}

We now apply a uniform thermal expansion, which induces 
the reactive axial force. 
This creates more compressive stress within the structure hence induces structural instability at a lower $u_p$ compared to the case with only mechanical loads. 
The optimized solutions for a range of $\alpha \Delta T$ at $u_p =0.08$ and $0.2$ are depicted in Fig.~\ref{fig:Res2_therm}. 
The 6-hole initial configuration is employed herein.
As noted, the temperature increase is shown to have a similar effect on the material layout as an increased force would have. 
With an increased temperature, the material layout tends to have a higher second moment of area and an increasing number of struts that reinforces the buckling capacity. 

\begin{figure}[hbt!]
    \centering
    \includegraphics[width=10cm]{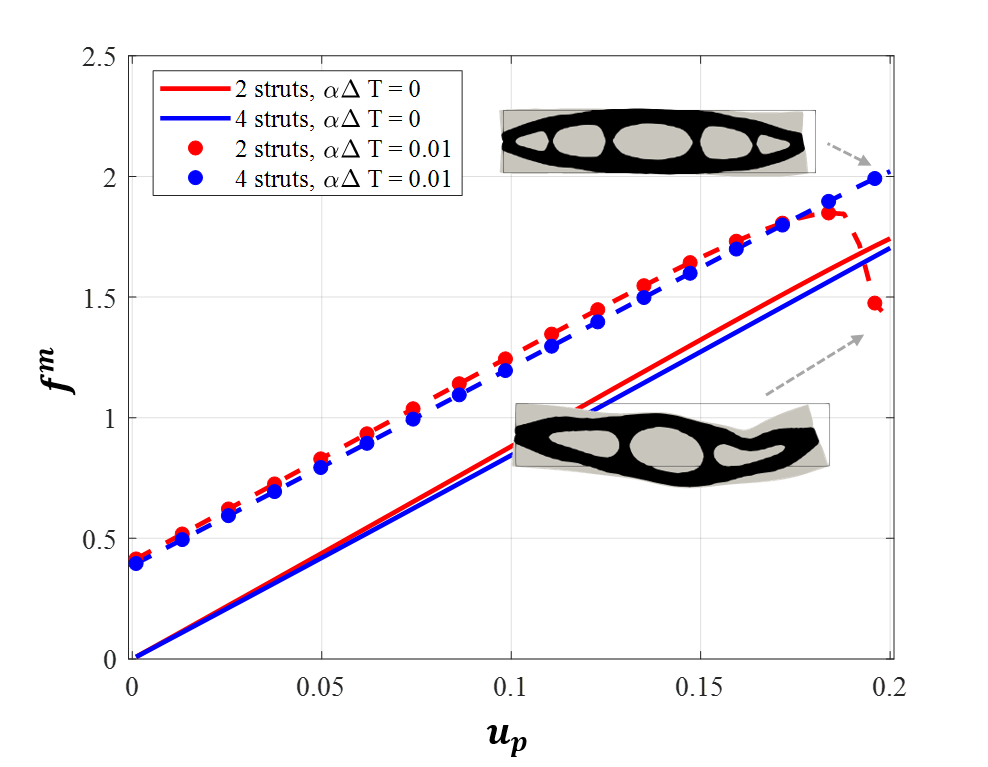}
    \caption{Structural responses of the two layouts that have different topology. In contrast to the mechanically-driven linear relation between force and displacement (solid lines), buckling is observed in the elevated temperature, making the 4-strut configuration preferred in the presence of the temperature change, as it restrains local buckling.}
    \label{fig:Res2_therm_FE}
\end{figure}

Figure~\ref{fig:Res2_therm_FE} illustrates how the increasing number of struts achieves an increased thermally-induced buckling capacity. 
We present the force-displacement curves for the optimum layouts found with $\alpha \Delta T = 10^{-2}$ (4 struts) and $\alpha \Delta T = 4 \times 10^{-3}$ (2 struts) and $u_p = 0.2$. 
The mechanical load $\boldsymbol f^m$ of these two representative layouts are computed by increasing $u_p$ from 0.0 to 0.2, for two temperature conditions: $\alpha \Delta T = 0.0 \text{ and } 0.01$.

It is observed that the 2-strut solution has slightly higher stiffness in the linear region for both $\alpha \Delta T =0$ and $0.01$.
This is because 2-strut topologies have more material along the longitudinal chords of the structure, which are the primary load-carrying passage in compression. 
When $u_p \geq 0.18$ and $\alpha \Delta T =0.01$, a local instability is induced by the thermal load as shown in the inset of Fig.~\ref{fig:Res2_therm_FE}.
It is clear from this illustration that optimization added the struts to prevent this local buckling. 

The capability to optimize a topology while considering buckling induced by thermal expansion is clearly demonstrated.
This opens up a variety of practical applications where buckling under high temperature needs to be suppressed.

\subsection{Short cantilever beam}
\begin{figure}[hbt!]
    \centering
    \includegraphics[width=10cm]{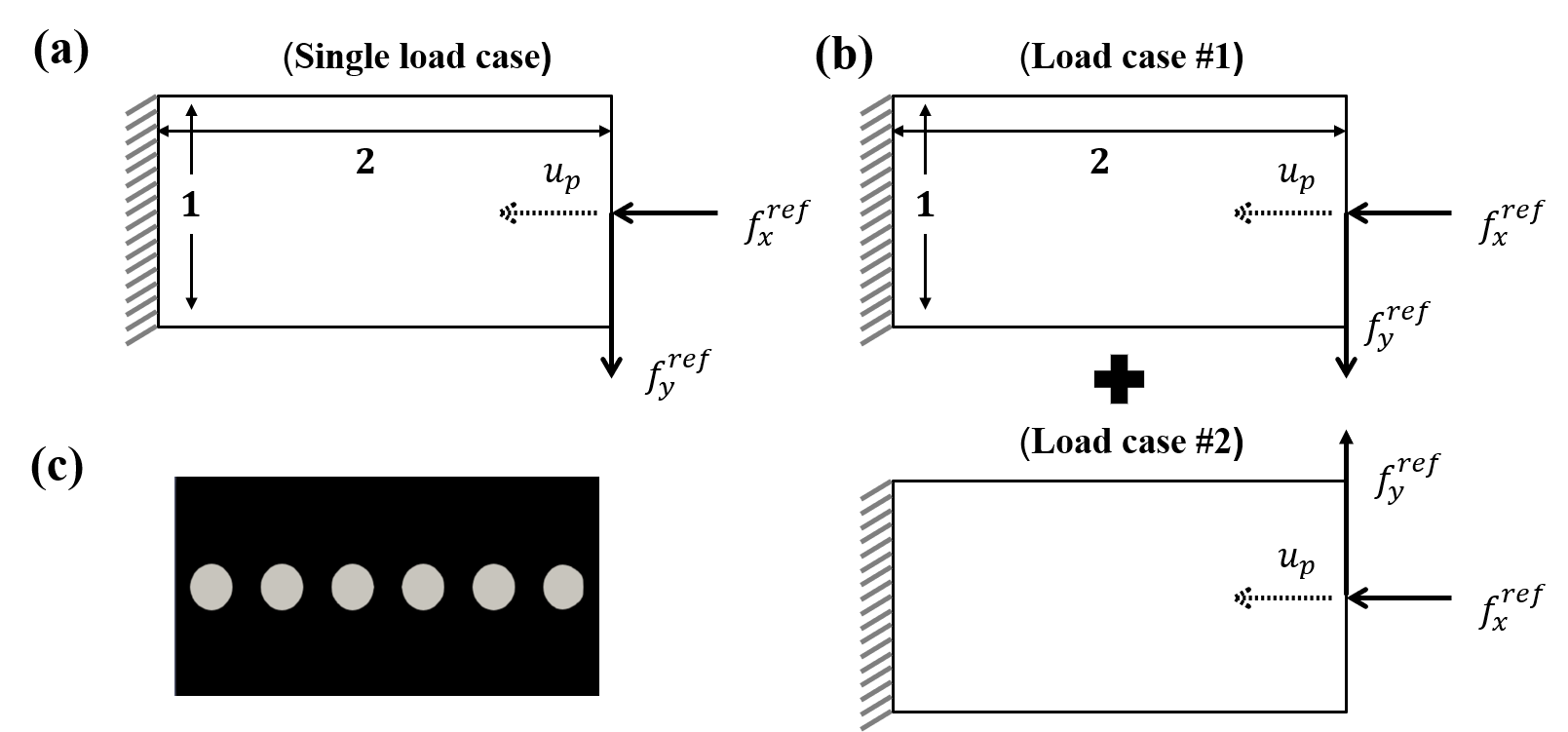}
    \caption{Design configuration of the cantilever beam example. (a) dimensions and boundary conditions of a single load case  problem; (b) dimensions and boundary conditions of a multiple load case problem; (c) initial material layouts, where voids are marked as gray. 
    }
    \label{fig:R1_ex4_config}
\end{figure}

A short cantilever is considered as shown in Fig.~\ref{fig:R1_ex4_config}. 
A controlled displacement $u_p$ is specified at the center node on the right boundary and constrains a nodal displacement in the x-direction. 
In contrast to the example with the uniaxially compressed load with a small eccentricity, the present example considers multiple loads composed of one horizontal load along the centerline $\hat{\boldsymbol{f}}_x^{ref}$, and a vertical load $\hat{\boldsymbol{f}}_y^{ref}$ at the same node.
The same load multiplication factor $\theta$ is used for these reference loads, although their magnitudes are treated independent of each other. 
The selection of the ratio between the loads (i.e., $\hat{\boldsymbol{f}}_y^{ref}/\hat{\boldsymbol{f}}_x^{ref}$) in part influences the optimization because it alternates the dominant deformation between in-plane compression to bending. 
We consider two cases, one as a single load case where the horizontal and vertical loads are applied simultaneously,
Fig.~\ref{fig:R1_ex4_config}(a) and the other as two load cases as shown in Fig.~\ref{fig:R1_ex4_config}(b). 
A uniform temperature change is applied to both.  
The initial material layout is shown in Fig.~\ref{fig:R1_ex4_config} (c), and rectangular design domain is discretized by 12,800 quadrilateral elements. 

\begin{figure}[hbt!]
    \centering
    \includegraphics[width=10cm]{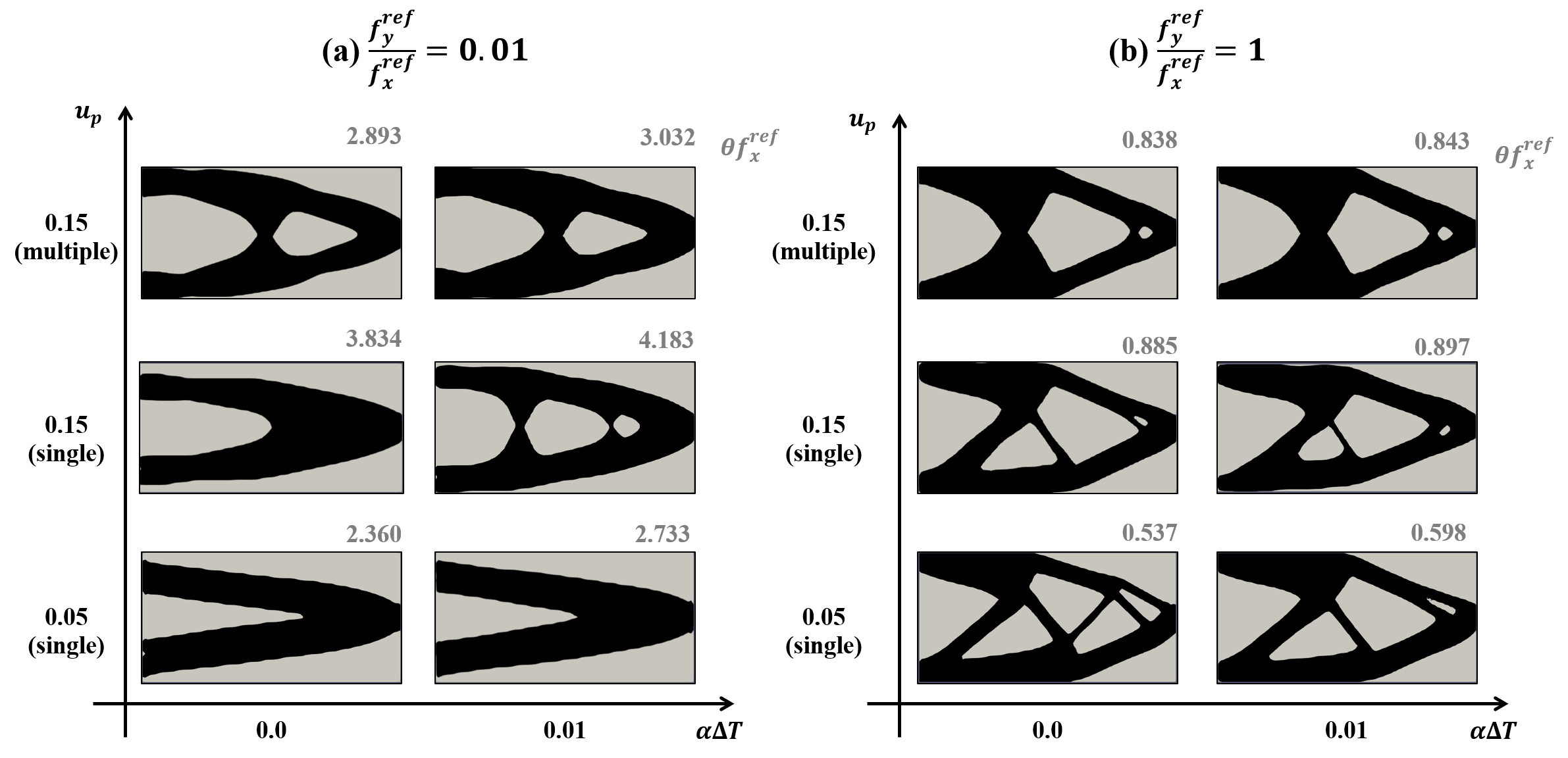}
    \caption{Optimum layouts for the cantilever beam. (a) $\hat{\boldsymbol{f}_y^{ref}}/\hat{\boldsymbol{f}_x^{ref}}=0.01$; (b) $\hat{\boldsymbol{f}_y^{ref}}/\hat{\boldsymbol{f}_x^{ref}}=1$. 
    }
    \label{fig:R1_ex_result}
\end{figure}

Two numerical values of $u_p$ are tested (0.05 and 0.15) for two different ratios of $\hat{\boldsymbol{f}_y^{ref}}/\hat{\boldsymbol{f}_x^{ref}}$, which are 0.01 and 1. 
Figure~\ref{fig:R1_ex_result} shows optimum material layouts for the set of problem configurations.
The optimum values of $\theta \hat{\boldsymbol{f}}_x^{ref}$ are given in gray on the top right of each solution. 

Figure~\ref{fig:R1_ex_result} shows the optimum structures for a range of configurations. 
First of all, the ratio between reference loads ($\hat{\boldsymbol{f}_y^{ref}}/\hat{\boldsymbol{f}_x^{ref}}$) is set to be 0.01 as shown in Fig~\ref{fig:R1_ex_result}(a). 
The structure deforms dominantly by in-plane contraction, as explicitly shown in small $u_p$. 
The designed layout optimized for $u_p=0.05$ shows that the horizontal compression is dominant with the additional small bending due to the vertical force. 
These optimum layouts also agree well with the layouts obtained via linear compliance minimization in which a cantilever with only central axial load is considered \cite{deng2017topology}. 
Adding a temperature does not change the optimum layout of $u_p=0.05$, although the required mechanical load $\theta\hat{\boldsymbol{f}_x^{ref}}$ increases by 15\% (to counteract the free thermal expansion). 

When $u_p$=0.15, the optimum topology changes to enhance the buckling rigidity of the structure by adding additional vertical reinforcements when heated ($\alpha\Delta T=0.01$).
This agrees well with Fig.~\ref{fig:Res2_mech} (a). 
For the multiple load case, symmetric material layouts are obtained although no symmetry is assumed in the design space. 
Symmetric designs, however, are shown to have a lower load-carrying capacity as shown in the decreased $\theta \hat{\boldsymbol{f}}^{ref}_x$ values when compared with the single load cases. 

When $\hat{\boldsymbol{f}_y^{ref}}/\hat{\boldsymbol{f}_x^{ref}}=1.0$, Fig.~\ref{fig:R1_ex_result}(b), 
The reference load toward vertical direction induces a significant bending. 
As a result, the optima found in $u_p=0.05$ and $u_p=0.15$ are better configured to sustain the bending. 
Since structure in bending loses a substantial in-plane load capacity, optimal $\theta \hat{\boldsymbol{f}_x^{ref}}$is substantially lower than for the equivalent in-plane, Fig.~\ref{fig:R1_ex_result}(a). 
Again, for the multiple load case, the symmetric design layouts are obtained.
The effect of the thermal load on the optimum layouts is not salient in the present example, 
because the boundary condition resembles a statically determinate beam.
Nevertheless, this example signifies the effect of considering simultaneous load and multiple load cases when displacement-controlled scheme is used. 

%% file: includes/Conclusion_new.tex
\section{Conclusion}

In this work, a level-set topology optimization formulation for nonlinear thermoelasticity is presented.
By introducing an intermediate state between the undeformed and deformed domains, the temperature-induced volume change is considered independent from the mechanical strain. 
Nonlinear finite element analysis and its consistent sensitivities are formulated based on a nonlinear strain measure along with multiplicative decomposition, hence the range of thermoelastic loads and design space are expanded in comparison to existing knowledge.

Four end-compliance minimization problems are employed to investigate the nonlinear thermoelastic structure design capabilities of the proposed formulation. 
The optimized layouts are observed to manipulate the thermal effect so that the given temperature change can counteract the mechanical load, even to the degree of suppressing buckling and nonlinear snap-through behavior. 
We construct a convex envelope of optimum layouts which represent the different ways that mechanical and thermal loads interact and influence the structural behavior. 
Such finding contrasts linear elastic optimization where the compliance is proportional to the temperature. 
The optimum results are shown to lie on the boundary of the envelope, which offer an effective means of understanding thermoelastic design solutions. 
Where a design is prone to buckle, the optimized structures for higher thermoelastic loads have material layouts with slits and struts, which in effect increase the buckling capacity. 

The careful investigations reveal that 
nonlinear elasticity is significant in thermoelastic topology optimization. 
The nonlinear coupled optimization method presented in this paper enables exploring a wider design space where 
the mechanical and thermal loads interact such a way that a lighter design compared to the mechanical or thermal load alone, may be present. 
We expect that more non-intuitive designs can be discovered with this optimization method.  

The present study suggests several new areas of future studies:
The present work limits the thermal expansion by up to 1\%. 
Although such range is well beyond the range that has been typically considered in the previous works,
a further study to investigate topology optimization for the thermal expansion greater than 1\% will offer a more in-depth understanding 
where thermally-induced nonlinearity becomes dominant.
The present study does not couple thermoelasticity with heat conduction, hence a temperature distribution is assumed to be known \textit{a priori}. 
However, this is a restrictive assumption as linear thermoelastic results considering design-dependent heat transfer demonstrated \cite{de2007topological, chen2010multiobjective}.
In addition, only a single type of the objective is considered in this work, where the end-compliance is minimized for a given mechanical and thermal loading conditions.
Other types of objective functions such as complementary energy which integrates the stored energy during the nonlinear loading path \cite{Buhl2000}, may need to be considered for a more comprehensive understanding.
An extension of the present 2D nonlinear thermoelastic topology optimization to 3D is the natural next step of this study, possibly tackled by using the state-of-the-art parallel computing \cite{aage2015topology} and an efficient data structure \cite{kambampati2018fast}.